\documentclass[iop]{emulateapj}
\usepackage{amsmath}

\begin{document}

\title{The Nature and Orbit of the Ophiuchus Stream}

\author{Branimir Sesar\altaffilmark{1,2}}
\author{Jo Bovy\altaffilmark{3,4}}
\author{Edouard J.~Bernard\altaffilmark{5}}
\author{Nelson Caldwell\altaffilmark{6}}
\author{Judith G.~Cohen\altaffilmark{7}}
\author{Morgan Fouesneau\altaffilmark{1}}
\author{Christian I.~Johnson\altaffilmark{6}}
\author{Melissa Ness\altaffilmark{1}}
\author{Annette M.~N.~Ferguson\altaffilmark{5}}
\author{Nicolas F.~Martin\altaffilmark{8,1}}
\author{Adrian M.~Price-Whelan\altaffilmark{9}}
\author{Hans-Walter Rix\altaffilmark{1}}
\author{Edward F.~Schlafly\altaffilmark{1}}
\author{William S.~Burgett\altaffilmark{13}}
\author{Kenneth C.~Chambers\altaffilmark{10}} 
\author{Heather Flewelling\altaffilmark{10}} 
\author{Klaus W.~Hodapp\altaffilmark{10}} 
\author{Nick Kaiser\altaffilmark{10}} 
\author{Eugene A.~Magnier\altaffilmark{10}}
\author{Imants Platais\altaffilmark{12}}
\author{John L.~Tonry\altaffilmark{10}}
\author{Christopher Waters\altaffilmark{10}}
\author{Rosemary~F.~G.~Wyse\altaffilmark{12}}

\altaffiltext{1}{Max Planck Institute for Astronomy, K\"{o}nigstuhl 17, D-69117
                 Heidelberg, Germany}
\altaffiltext{2}{Corresponding author: bsesar@mpia.de}
\altaffiltext{3}{Institute for Advanced Study, Einstein Drive, Princeton, NJ
                 08540, USA}
\altaffiltext{4}{John Bahcall Fellow}
\altaffiltext{5}{SUPA, Institute for Astronomy, University of Edinburgh, Royal
                 Observatory, Blackford Hill, Edinburgh EH9 3HJ, UK}
\altaffiltext{6}{Harvard-Smithsonian Center for Astrophysics, 60 Garden Street,
                 Cambridge, MA 02138, USA}
\altaffiltext{7}{Division of Physics, Mathematics and Astronomy, California
                 Institute of Technology, Pasadena, CA 91125,
                 USA}
\altaffiltext{8}{Observatoire astronomique de Strasbourg, Universit\'e de
                 Strasbourg, CNRS, UMR 7550, 11 rue de l'Universit\'e, F-67000
                 Strasbourg, France}
\altaffiltext{9}{Department of Astronomy, Columbia University, 550 West 120th Street, New York, NY 10027, USA}
\altaffiltext{10}{Institute for Astronomy, University of Hawaii at Manoa, Honolulu, HI 96822, USA}
\altaffiltext{11}{Department of Physics, Durham University, South Road, Durham DH1 3LE, UK}
\altaffiltext{12}{Department of Physics and Astronomy, Johns Hopkins University, 3400 North Charles Street, Baltimore, MD 21218, USA}
\altaffiltext{13}{GMTO Corporation, 251 S.~Lake Ave., Suite 300, Pasadena, CA
                  91101, USA } 

\begin{abstract}
The Ophiuchus stream is a recently discovered stellar tidal stream in the Milky
Way. We present high-quality spectroscopic data for 14 stream member stars
obtained using the Keck and MMT telescopes. We confirm the stream as a fast
moving ($v_{los}\sim290$ km s$^{-1}$), kinematically cold group
($\sigma_{v_{los}}\lesssim1$ km s$^{-1}$) of $\alpha$-enhanced and metal-poor
stars (${\rm [\alpha/Fe]\sim0.4}$ dex, ${\rm [Fe/H]\sim-2.0}$ dex). Using a
probabilistic technique, we model the stream simultaneously in line-of-sight
velocity, color-magnitude, coordinate, and proper motion space, and so determine
its distribution in 6D phase-space. We find that that the stream extends in
distance from 7.5 to 9 kpc from the Sun; it is 50 times longer than wide, merely
appearing highly foreshortened in projection. The analysis of the stellar
population contained in the stream suggests that it is $\sim12$ Gyr old, and
that its initial stellar mass was $\sim2\times10^4$ $M_\sun$ (or at least
$\ga7\times10^3$ $M_\sun$). Assuming a fiducial Milky Way potential, we fit an
orbit to the stream which matches the observed phase-space distribution, except
for some tension in the proper motions: the stream has an orbital period of
$\sim350$ Myr, and is on a fairly eccentric orbit ($e\sim0.66$) with a
pericenter of $\sim3.5$ kpc and an apocenter of $\sim17$ kpc. The phase-space
structure and stellar population of the stream show that its progenitor must
have been a globular cluster that was disrupted only $\sim240$ Myr ago. We do
not detect any significant overdensity of stars along the stream that would
indicate the presence of a progenitor, and conclude that the stream is all that
is left of the progenitor.
\end{abstract}


\section{Introduction}\label{introduction}

One of the main goals of Galactic astronomy is the measurement of the Milky
Way's gravitational potential, because knowledge of it is required in any study
of the dynamics or evolution of the Galaxy. An important tool in this
undertaking are stellar tidal streams, remnants of accreted Milky Way satellites
that were disrupted by tidal forces and stretched into filaments as they orbited
in the Galaxy's potential. The orbit of a stream is sensitive to the properties
of the potential and thus can be used to constrain the potential over the range
of distances spanned by the stream (e.g., \citealt{kop10, new10, ses13, bel14}).
In this context, the recently discovered Ophiuchus stellar stream \citep{ber14b}
is very interesting because it is located fairly close to the Galactic center
(galactocentric distance of $\sim5$ kpc), and as such probes the part of the
potential that other known stellar tidal streams do not probe.

The Ophiuchus stream is a $\sim2.5\arcdeg$ long and $7\arcmin$ wide stellar
stream that was recently discovered by \citet{ber14b} in the Pan-STARRS1
photometric catalog (PS1; \citealt{kai10}). Bernard et al.~inferred from its
color-magnitude diagram (CMD) that it is consistent with an old ($\ga10$ Gyr)
and relatively metal-poor population ($[Fe/H]\sim -1.3$ dex) located $\sim9$ kpc
away at $(l, b)\sim(5\arcdeg, +32\arcdeg)$. They did not detect a progenitor (or
a remnant of it), but suggested that the progenitor would most likely be a
globular cluster.

Due to the lack of proper motion and line ofsight velocity measurements,
Bernard et al.~could not determine the orbit of stream and thus could not use it
to constrain the potential. Furthermore, without knowing the orbit of the
stream, they could not fully explain two interesting properties of the
Ophiuchus stream, namely, its very short length and the lack of a visible
progenitor. The projected angular length of $2.5\arcdeg$ at a distance of
$\sim9$ kpc implies a projected physical length of $\sim400$ pc for the
Ophiuchus stream. Such a short length suggests that its progenitor must have
been disrupted fairly recently. However, if that was the case, the progenitor
should still be visible as an overdensity of stars somewhere along the stream.
Yet, no progenitor has been detected so far.

To address the above questions, we need to know the orbit of the Ophiuchus
stream, and to determine its orbit we need to measure the stream's line of sight
velocity, distance, and proper motion. In Section~\ref{data}, we describe the
data we use in this work; the PS1 photometry and astrometry, the spectroscopic
follow-up of candidate stream members, and the measurement of their
line of sight velocities, element abundances, and proper motions. In
Section~\ref{results}, we provide a detailed characterization of the stream in
position, velocity, and abundance (7D) phase space. The constraints obtained in
Section~\ref{results} are then used to constrain and examine the orbit of the
stream (Section~\ref{orbit}) and the time of disruption of its progenitor
(Section~\ref{disruption}). In Section~\ref{conclusions}, we discuss the nature
of the stream's peculiar orbit, highlight the solved and uncovered puzzles
related to the stream, and finally present our conclusions.

\section{Data}\label{data}

\subsection{Overview of the Pan-STARRS1 survey}\label{PS1}

The PS1 survey has observed the entire sky north of declination $-30\arcdeg$ in
five filters covering $400-1000$ nm \citep{stu10, ton12}. The 1.8-m PS1
telescope has a 7 deg$^2$ field of view outfitted with a billion-pixel camera
\citep{hod04, ona08, to09}. In single-epoch images, the telescope can detect
point sources at a signal-to-noise ratio (S/N) of 5 at 22.0, 22.0, 21.9, 21.0,
and 19.8 mag in PS1 $grizy_{P1}$ bands, respectively. The survey pipeline
automatically processes images and performs photometry and astrometry on
detected sources \citep{mag06, mag07}. The uncertainty in photometric
calibration of the survey is $\la0.01$ mag \citep{sch12}, and the astrometric
precision of single-epoch detections is 10 milliarcsec
\citep[hereafter mas]{mag08}.

\subsection{Line of sight velocities}

Based on the findings of \citet{ber14b}, we have used the dereddened fiducial of
the old globular cluster NGC 5904 (from \citealt{ber14a}, and shifted to the
distance modulus of 14.9 mag) to select $\sim170$ candidate Ophiuchus stream
members from the PS1 photometric catalog. The candidates were selected if their
position was within $4.5\arcmin$ of the best-fitting great circle containing the
stream (see Figure 1 of \citealt{ber14b}), and if their dereddened
$g_{P1}-i_{P1}$ color and $i_{P1}$-band magnitude were within 0.1 mag and 0.5
mag of the fiducial isochrone, respectively. We observed the selected candidates
using the DEIMOS spectrograph on Keck II \citep{fab02} and using the Hectochelle
fiber spectrograph on MMT \citep{sze11} over a course of two nights.

Seven candidate blue horizontal branch stars were observed with DEIMOS on 2014
May 29th (project ID 2014A-C171D, PI: J.~Cohen). The observations were made
using the $0.8\arcsec$ slit and the high resolution (1200 G) grating, delivering
a resolution of 1.2 {\AA} in the 6250-8900 {\AA} range. The spectra were
extracted and calibrated using standard
IRAF\footnote{\url{http://iraf.noao.edu/}} tasks. The uncertainty in the
zero-point of wavelength calibration (measured using sky lines) was
$\lesssim0.04$ {\AA} ($\lesssim2$ km s$^{-1}$ at 6563 \AA).

The line of sight velocities of stars observed by DEIMOS were measured by
fitting observed spectra with synthetic template spectra selected from the
\citet{mun05} spectral
library\footnote{\url{http://archives.pd.astro.it/2500-10500/}}. Prior to
fitting, the synthetic spectra were resampled to the same \AA~per pixel scale
as the observed spectrum and convolved with an appropriate Line Spread Function.
The velocity obtained from the best-fit template was corrected to the
barycentric system and adopted as the line of sight velocity, $v_{\rm los}$. We
added in quadrature the uncertainty in the zero-point of wavelength calibration
(2 km s$^{-1}$ at 6563 \AA) to the velocity error from fitting.

The remaining 163 Ophiuchus stream candidates were observed with Hectochelle on
2014 June 6th (proposal ID 2014B-SAO-4, PI: C.~Johnson). Observations were made
using the RV31 radial velocity filter, which includes Mg I/Mgb features in the
5150-5300 {\AA} range. To improve the S/N of faint targets, we binned the
detector by 3 pixels in the spectral direction, resulting in an effective
resolution of $R\sim38,000$.

Hectochelle spectra were extracted and calibrated following \citet{cal09}. To
account for variations in the fiber throughput, the spectra were normalized
before sky subtraction. The normalization factor was estimated using the
strength of several night sky emission lines in the appropriate order. Sky
subtraction was performed using the average of 20-30 sky fibers, using the
method devised by \citet{kop11}. A comparison of observed and laboratory
positions of sky emission lines did not reveal any significant offsets in
wavelength calibration (i.e., no offsets greater than 0.5 km s$^{-1}$ at 5225
\AA).

The line of sight velocities of stars observed by Hectochelle were measured
using the RVSAO package \citep{km98}, by cross-correlating observed spectra with
a synthetic spectrum of an A-type and a G-type giant star (constructed by
\citealt{lat02}). The velocity obtained from the best fitting template was
adopted. To the uncertainty in $v_{\rm los}$, measured by RVSAO, we added (in
quadrature) the uncertainty in the zero-point of wavelength calibration, which
we measured using sky emission lines to be 0.5 km s$^{-1}$. Finally, the
measured velocities were corrected to the barycentric system using the BCVCORR
task.

A comparison of velocities measured from DEIMOS and Hectochelle spectra for star
``bhb6'' (Table~\ref{table1}), shows that the two velocity sets are consistent
within stated uncertainties.

\subsection{Element abundances}

Even though the primary goal of spectroscopic observations was to obtain precise
radial velocities, the wavelength range and the resolution of Hectochelle
spectra are sufficient to allow estimates of element abundances.

We determined stellar parameters from the continuum-normalized, radial
velocity-corrected spectra using the SMH code of \citet{cas14}, which is built
on the MOOG code of \citet{sne73}. Kurucz $\alpha-$enhanced (0.4 dex) model
atmospheres \citep{ck04} and a line list compiled from \citet{fre10} by
\citet{cas14} were used (see Table~5 in the electronic version of the Journal).
First, effective temperatures were calculated from 2MASS photometry and
color-temperature calibrations of \citet{ghb09} and spectroscopic temperatures
were optimized around this value using the SMH code, by removing abundance
trends with line excitation potential. Parameters of $\log g$ and ${\rm [Fe/H]}$
were determined using 14 Fe I and 2-3 Fe II lines to achieve ionization balance,
and microturbulence was calculated by removing trends in abundances as a
function of the reduced equivalent width of the lines. Estimates of the
$\alpha-$enhancement were obtained using only the few available clean Mg, Ca and
Ti (I and II) lines, which comprised a total of 6-8 lines per star. The
abundances were measured from equivalent widths and the lines we used are listed
in Table~5. Because Mg lines are strong and may be saturated, the values of
${\rm [Mg/Fe]}$ are significantly different than values of ${\rm [Ca/Fe]}$ and
${\rm [Ti/Fe]}$.

\subsection{Proper motions}\label{proper_motions}

Proper motions are crucial constraints for determining the orbit of a stream
(e.g., \citealt{kop10}). To measure the proper motion of stars in the vicinity
of the Ophiuchus stream, we combine the astrometry provided by USNO-B
\citep{mon03} and 2MASS \citep{skr06} catalogs with the PS1 catalog. The USNO-B 
catalog lists photometry and astrometry measured from photographic plates in
five different band-passes ($O$, $E$, $J$, $F$, and $N$). The USNO-B plates were
exposed at different epochs, and thus each object in the catalog can have a
maximum of five recorded positions. The 2MASS catalog provides only one position
entry per object.

To reduce the systematic offsets in astrometry between different catalogs, we
first calibrate USNO-B and 2MASS positions to a reference frame that is defined
by positions of galaxies observed in PS1 (which are on the ICRS coordinate
system). We define galaxies as objects that have the difference between
point-spread function (PSF) and aperture magnitudes in PS1 $r_{P1}$ and $i_{P1}$
bands between 0.3 and 1.0 mag.

The astrometric reference catalog is created by averaging out repeatedly
observed positions of PS1 objects. Between 2012 May and June, the region in the
vicinity of the Ophiuchus stream was observed four times in PS1 $g_{P1}$,
$r_{P1}$, and $i_{P1}$ bands. To minimize the uncertainty in astrometry due to
wavelength-dependent effects, such as the differential chromatic refraction, we
only average out positions observed through the $r_{P1}$-band filter. Since the
astrometric precision of single-epoch detections is 10 mas \citep{mag08}, the
precision of the average position is $\sim5$ mas or better.

The USNO-B astrometry is calibrated following
\citet[see their Section 2.1]{mun04}. First, we calculate the positions
of objects at each of the five USNO-B epochs, using software kindly provided by
J.~Munn. Then, for each USNO-B object we find the 100 nearest galaxies,
calculate the median offsets in right ascension and declination between the
reference PS1 position and the USNO-B position for these galaxies, and add the
offsets to the USNO-B position in question. This is done separately for each of 
the five USNO-B epochs. The single-epoch 2MASS positions are calibrated using
the same procedure.

Having tied the positions for each object at one 2MASS and five USNO-B epochs to
the PS1 astrometric reference frame, we can now check for any additional
systematic uncertainties in the calibrated astrometry. We do so using the
leave-one-out cross-validation. One of the six calibrated positions is withheld,
and a straight line is fitted to the remaining five positions and the PS1
position. The straight line fit (i.e., essentially a proper motion fit,
neglecting the parallax) is then used to predict the position of an object at
the withheld epoch. The difference between the withheld position and the
predicted position is labeled as $\Delta RA$ or $\Delta Decl$.

\begin{figure}
\plotone{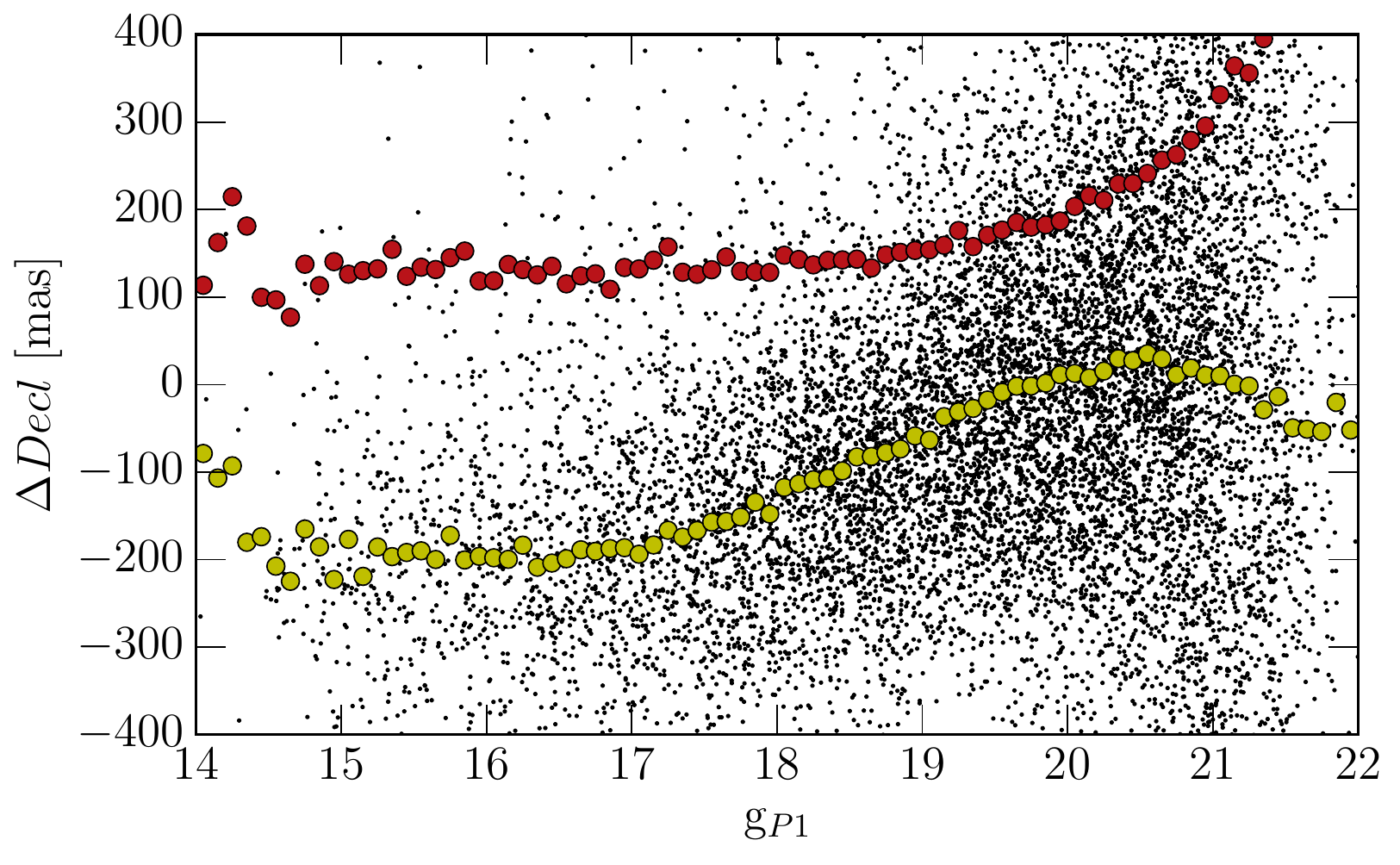}
\caption{
This plot illustrates the systematic offset in declination ($\Delta Decl$) of
objects observed in the POSS-II Blue epoch of the USNO-B catalog (plate 799), as
a function of the $g_{P1}$-band magnitude. For clarity, only a subset of objects
are plotted. The solid yellow circles show the median $\Delta Decl$ in magnitude
bins, and the solid red circles show the rms scatter in magnitude bins. Note how
the brighter objects are systematically offset by $\sim200$ mas from the fainter
objects. The rms scatter indicates that the average precision in this coordinate
and epoch is $\sim120$ mas for objects brighter than $g_{P1}=19$.
\label{usnob_crossvalidation}}
\end{figure}

Inspection of $\Delta RA$ or $\Delta Decl$ values has revealed that the
positions of USNO-B objects depend on magnitude for some epochs
(Figure~\ref{usnob_crossvalidation}). We have examined $\Delta RA$ and
$\Delta Decl$ values in different regions of the sky, and have concluded that
these astrometric issues affect individual photographic plates, and are not
specific to a particular photographic bandpass. To remove this dependence, we
subtract plate-specific and magnitude-dependent offsets (shown by yellow circles
in Figure~\ref{usnob_crossvalidation}) from the original USNO-B positions
{\em before} we calibrate the positions using PS1 galaxies. Since USNO-B does
not provide uncertainty in positions, we adopt the rms scatter of $\Delta RA$
and $\Delta Decl$ values (shown by red solid circles in
Figure~\ref{usnob_crossvalidation}) as an estimate of the uncertainty in
position at a given magnitude and epoch.

We have also examined whether $\Delta RA$ and $\Delta Decl$ astrometric
residuals depend on the $g_{P1}-i_{P1}$ color. We find that the residuals do
depend on the color, and that they can be as high as 100 mas. The most likely
explanation for this dependence is the differential chromatic refracation. We
correct for this dependence using a similar approach as above. For each
photographic plate we bin $\Delta RA$ and $\Delta Decl$ residuals as a function
of color, calculate the median for each color bin, and subtract that value from
the observed positions of stars in that color bin, {\em before} we calibrate the
positions using PS1 galaxies.

Finally, to measure the absolute proper motion of an object we fit a straight
line to all available positions (min.~3, max.~7) as a function of time (the
epoch baseline is 58 years). The proper motion of confirmed Ophiuchus stream
members is listed in Table~\ref{table1}.

For verification, we have also measured the proper motion of 700 candidate QSOs,
selected using $WISE$ \citep{wri10} $W1-W2>0.8$ color criterion (see Section 2.1
by \citealt{nik14}). The median proper motion of candidate QSOs is 0.3 mas
yr$^{-1}$ and the uncertainty of the median is 0.2 mas yr$^{-1}$, showing that
there is no statistically significant offset in measured proper motions.

To measure the systematic uncertainty in proper motions, one should ideally
calculate the rms scatter of proper motions of fairly bright and static sources.
Bright QSOs would be an ideal choice for this measurement, because they are
extragalactic point sources and because they are not used in the calibration
process. Unfortunately, the candidate QSOs described above are too faint ($r>18$
mag) to be used for this purpose (i.e., the uncertainty in their proper motions 
is already dominated by Poisson noise).

Instead, we measure the systematic uncertainty by calculating the median
uncertainty in proper motion of bright stars. For stars brighter than $r=17$
mag, the uncertainty in proper motion is $\sim1.5$ mas/yr (and constant with
magnitude), and we adopt this value as the systematic uncertainty. To determine
whether this uncertainty is well-measured, we examined the distribution of
$\chi^2$ per-degrees of freedom values calculated from proper motion fits of
bright stars ($r < 17$). The mode of this distribution is centered at 1,
indicating that on average, the uncertainties in proper motion are well-measured
(i.e., not overestimated or underestimated). However, the observed distribution 
has longer tails than expected (toward high $\chi^2$ values), indicating that
for some stars the uncertainties in proper motion are understimated (e.g., due
to blending of sources in photographic plates).

\section{Characterization of the Ophiuchus stream}\label{results}

\subsection{Line of sight velocities\label{vlos}}

\begin{figure}
\plotone{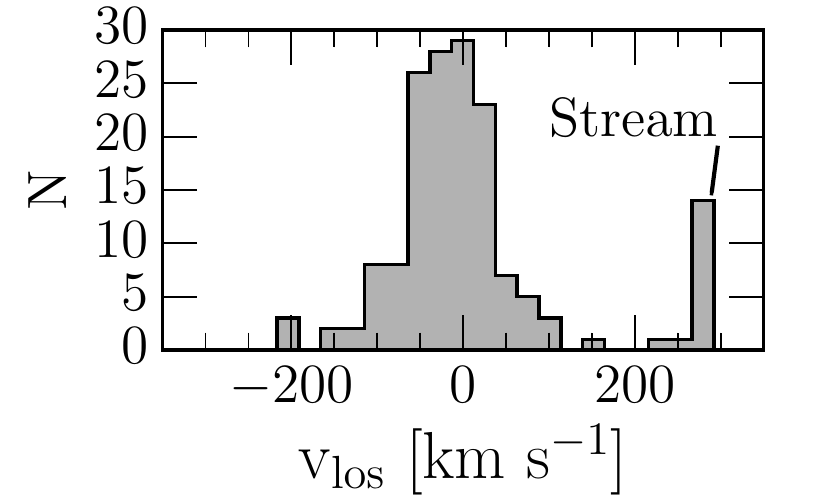}
\caption{
The distribution of heliocentric line of sight velocities of stars observed by
DEIMOS and Hectochelle. The uncertainty in individual $v_{los}$ measurements is
$\lesssim2$ km s$^{-1}$ and the bin size is 25 km s$^{-1}$. The Ophiuchus stream
is detected as a group of stars with $v_{los}\sim290$ km s$^{-1}$.
\label{rv_hist}}
\end{figure}

The $v_{los}$ distribution of stars observed by DEIMOS and Hectochelle is shown
in Figure~\ref{rv_hist}. In this figure, a group of 14 stars with
$285 < v_{los}/ {\rm km s^{-1}} < 292$ clearly stands out. This group, which we
identify as the Ophiuchus stream, is well-separated from the majority of stars
which have $|v_{los}|<200$ km s$^{-1}$. The positions, velocities, and PS1
photometry of stars in this group are listed in Table~\ref{table1}.

\begin{figure}
\plotone{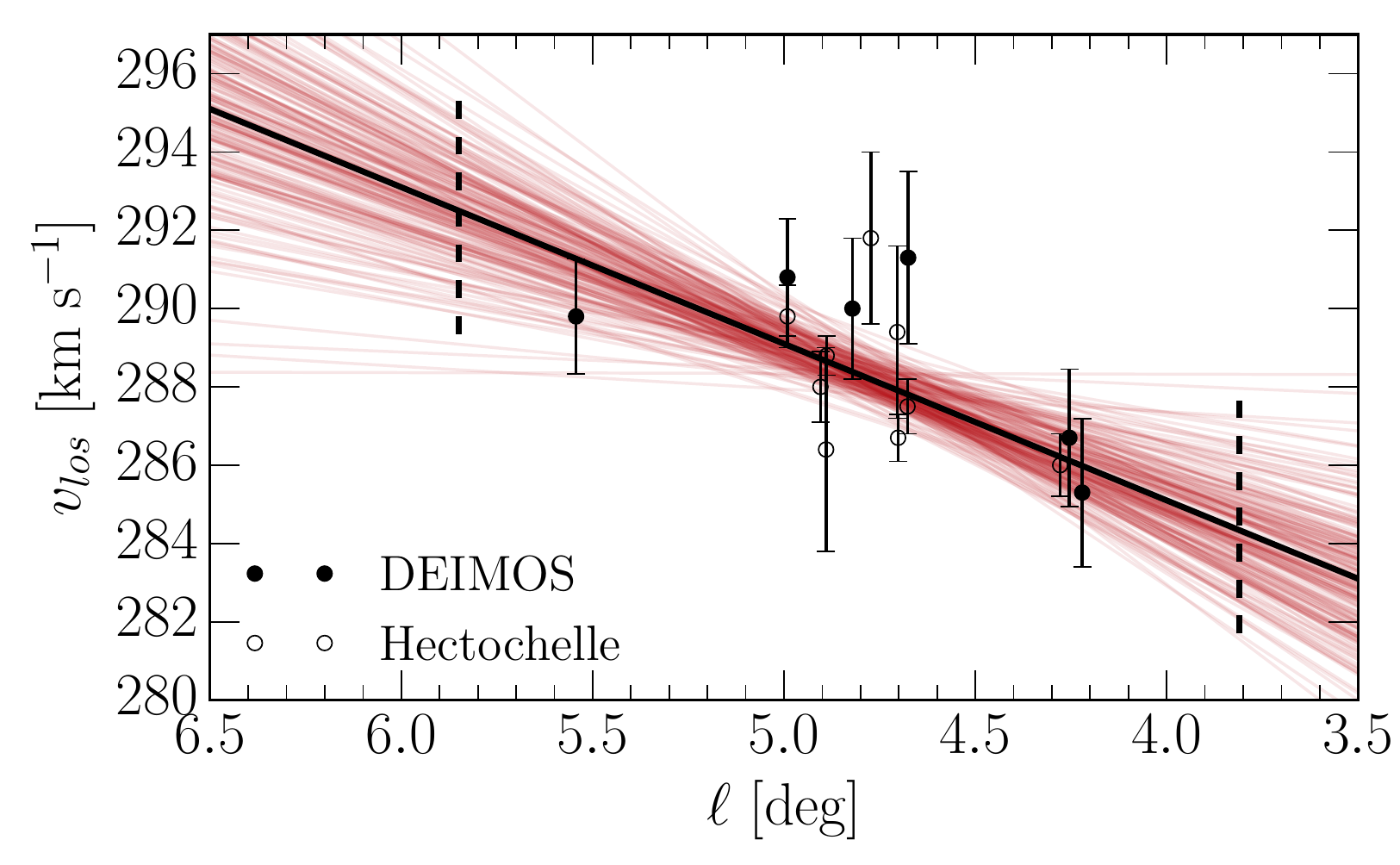}
\caption{
Line-of-sight velocities of stars in the Ophiuchus stream are shown as symbols
with error bars. The thick solid line shows the most probable model
($v_{los}(\ell)=4.0(\ell-5)+289.1$ km s$^{-1}$). To illustrate the uncertainty
in the most probable model, the thin semi-transparent red lines show 200 models
sampled from the posterior distribution. The vertical dashed lines show the
likely extent of the stream (see Section~\ref{proper_motion_extent}).
\label{gl_vs_rv}}
\end{figure}

A closer look at $v_{los}$ of stars in the Ophiuchus stream
(Figure~\ref{gl_vs_rv}) suggests that their velocities are changing as a
function of galactic longitude. To fit this possible velocity gradient, we use
an approach similar to the one taken by \citet[see their Section 2.1.1]{mj10}.

We wish to find the most likely set of parameters $\theta$ for which the
observations of stars listed in Table~\ref{table1},
$\mathcal{D}={\bf \{d_k\}}_{1\leq k\leq14}$, match the model described below. In
the current problem, each data point ${\bf d_k}$ is defined by its line of sight
velocity $v_{los,k}$ at galactic longitude $\ell_k$, $d_k=\{v_{los,k},\ell_k\}$.
The velocity has an uncertainty of $\sigma_{v_{los,k}}$. The star ``bhb6'' has
been observed twice, so for this star we adopt the weighted average (and its
associated uncertainty) of the two line of sight velocity observations. The
uncertainty in longitude is not considered because it is much smaller compared
to the uncertainty in velocity. The data points are also considered to be
independent. Therefore, the likelihood of these data points with the model
defined by the set of parameters $\theta$, is
\begin{equation}
p(\mathcal{D} | \theta) = \prod_k p({\bf d_k | \theta, \ell_k})\label{total_likelihood},
\end{equation}
where $p({\bf d_k | \theta, \ell_k})$ is the likelihood of data point $k$ to be
generated from the model. Using Bayes theorem, the probability of a model
given the data, $p(\theta | \mathcal{D})$, is
\begin{equation}
p(\theta | \mathcal{D}) \propto p(\mathcal{D} | \theta)p(\theta)\label{probability_of_a_model},
\end{equation}
where $p(\theta)$ represents our prior knowledge on the model.

We explicitly define the likelihood $p({\bf d_k | \theta, \ell_k})$ as
\begin{equation}
 p(v_{los,k} | \theta, \sigma_{v_{los, k}}, \ell_k) = \mathcal{N}(v_{los,k} | v(\ell_k), \sigma^\prime_k)\label{vel_likelihood},
\end{equation}
where
\begin{equation}
\mathcal{N}(x|\mu,\sigma)=(1/\sqrt{2\pi\sigma^2})\exp(-0.5((x-\mu)^2)/\sigma^2)
\end{equation} is a normal distribution, and
$\theta=\{\frac{dv_{los}}{d\ell}, \overline{v_{los}}, s\}$ are parameters of our
model:
\begin{enumerate}
    \item $\frac{dv_{los}}{d\ell}$ is the velocity gradient along the galactic
        longitude direction,
    \item $\overline{v_{los}}$ is the velocity of the stream at the reference
        galactic longitude $\ell_0=5\arcdeg$, and
    \item $s$ accounts for the additional scatter in velocities (e.g., due to
        the intrinsic velocity dispersion of the stream).
\end{enumerate}
The $v(\ell_k)$ is the predicted velocity of the stream at position $\ell_k$
\begin{equation}
v(\ell_k) = \frac{dv_{los}}{d\ell}(\ell_k - \ell_0) +\overline{v_{los}}
\end{equation}
and $\sigma^\prime_k=\sqrt{s^2 + \sigma^2_{v_{los, k}}}$ is the quadratic sum of
the additional scatter $s$ in velocity and the uncertainty in line of sight
velocity of data point $k$. The likelihood of all data points can be calculated
using Equation~\ref{total_likelihood}.

Before we can sample from the posterior probability distribution over our model
parameters, we need to define the prior probabilities of model parameters. As
prior probabilities, we adopt priors that are uniform in these ranges:
$270 < \overline{v_{los}}/{\rm km\, s^{-1}} < 320$,
$-8 < \frac{dv_{los}}{d\ell}/{\rm km\, s^{-1}\, deg^{-1}} < 8$,
$0 \leq s/{\rm km\, s^{-1}} < 3$.

To efficiently explore the parameter space, we use the \citet{gw10} Affine
Invariant Markov chain Monte Carlo (MCMC) Ensemble sampler as implemented in the
\texttt{emcee} package\footnote{\url{http://dan.iel.fm/emcee/current/}} (v2.1,
\citealt{fm12}). We use 200 walkers and obtain convergence\footnote{We checked
for convergence of chains by examining the auto-correlation time of the chains
per dimension.} after a short burn-in phase of 100 steps per walker. The chains
are then restarted around the best-fit value and evolved for another 2000 steps.
To enable easy reconstruction of the posterior distribution, we provide chains
in a Zenodo data repository\footnote{\url{https://zenodo.org/record/19197}}
\citep[doi:10.5281/zenodo.19197]{Sesar2015data} for all of the data modeling in
Section~\ref{results} of this paper.

We characterize the most probable set of parameters for our model using the
maximum a posteriori values. We also report the median and equivalent 1-$\sigma$
confidence intervals using the 50th, 16th and 84th percentiles, respectively
(see Table~\ref{stream_parameters}).

We find $\frac{dv_{los}}{d\ell}=4.0\pm1.2$ km s$^{-1}$ deg$^{-1}$,
$\overline{v_{los}}=289.1\pm0.4$ km s$^{-1}$, and a very small velocity
dispersion of $s=0.4_{-0.4}^{+0.5}$ km s$^{-1}$. Thus, we detect a gradient in
the line of sight velocity at a $4\sigma$ level.

\begin{deluxetable}{lrr}
\tabletypesize{\scriptsize}
\setlength{\tabcolsep}{0.02in}
\tablecolumns{3}
\tablewidth{0pc}
\tablecaption{Ophiuchus stream parameters\label{stream_parameters}}
\tablehead{
\colhead{Parameter} & \colhead{MAP$^a$} & \colhead{Median and central $68\%$ C.I.$^b$}
}
\startdata
$\overline{v_{los}}$ & 289.1 & $289.1_{-0.4}^{+0.4}$ km s$^{-1}$ \\
$\frac{dv_{los}}{d\ell}$ & 4.0 & $4.0_{-1.2}^{+1.2}$ km s$^{-1}$ deg$^{-1}$ \\
$s$ & 0.0 & $0.4_{-0.4}^{+0.5}$ km s$^{-1}$ \\
${\rm [Fe/H]}$ & & $-1.95_{-0.5}^{+0.5}$ dex \\
${\rm [\alpha/Fe]}$ & & $0.4_{-0.1}^{+0.1}$ dex \\
\hline \\
Age & 11.7 & $11.7_{-0.3}^{+0.6}$ Gyr \\
Mass-loss parameter $\eta$ & 0.49 & $0.48_{-0.04}^{+0.02}$ \\
Metallicity $Z$ & $2.3\times 10^{-4}$ & $(2.3_{-0.3}^{+0.4})\times 10^{-4}$ \\
$E(B-V)_{off}^c$ & 0.011 & $0.008_{-0.009}^{+0.009}$ mag \\
$\overline{DM}$ & 14.57 & $14.58_{-0.05}^{+0.05}$ mag \\
$\frac{dDM}{d\ell}$ & -0.20 & $-0.20_{-0.03}^{+0.03}$ mag deg$^{-1}$ \\
$\sigma_{g,iso}$ & 0.0003 & $0.012_{-0.008}^{+0.016}$ mag \\
$\sigma_{r,iso}$ & 0.0003 & $0.009_{-0.006}^{+0.012}$ mag \\
$\sigma_{i,iso}$ & 0.0006 & $0.007_{-0.005}^{+0.009}$ mag \\
$\sigma_{z,iso}$ & 0.0003 & $0.006_{-0.005}^{+0.008}$ mag \\
$\sigma_{y,iso}$ & 0.0003 & $0.007_{-0.005}^{+0.009}$ mag \\
\hline \\
$\ell_{min}$ & 3.81 & $3.84_{-0.03}^{+0.03}$ deg \\
$\ell_{max}$ & 5.85 & $5.86_{-0.03}^{+0.03}$ deg \\
A & 31.37 & $31.38_{-0.02}^{+0.02}$ deg \\
B & -0.80 & $-0.80_{-0.03}^{+0.03}$ \\
C & -0.15 & $-0.16_{-0.04}^{+0.04}$ deg$^{-1}$ \\
Deprojected length & 1.6 & $1.5_{-0.3}^{+0.3}$ kpc \\
$\sigma_b$ & 6.0 & $6.9_{-0.6}^{+0.7}$ arcmin \\
$\overline{\mu_\ell}$ & -5.5 & $-5.6_{-0.3}^{+0.3}$ mas yr$^{-1}$ \\
$\frac{d\mu_\ell}{dl}$ & -1.5 & $-1.6_{-0.6}^{+0.5}$ mas yr$^{-1}$ deg$^{-1}$ \\
$\overline{\mu_b}$ & 2.4 & $2.3_{-0.3}^{+0.3}$ mas yr$^{-1}$ \\
$\frac{d\mu_b}{d\ell}$ & 2.0 & $2.3_{-0.4}^{+0.5}$ mas yr$^{-1}$ deg$^{-1}$ \\
\hline \\
Pericenter & 3.55 & $3.57_{-0.06}^{+0.05}\left(_{-0.05}^{+0.35}\right)^d$ kpc \\
Apocenter & 17.0 & $16.8_{-0.4}^{+0.6}\left(_{-2.9}^{+0.0}\right)$ kpc \\
Eccentricity & 0.66 & $0.65_{-0.01}^{+0.01}\left(_{-0.08}^{+0.0}\right)$ \\
Orbital period & 351 & $346_{-7}^{+11}\left(_{-73}^{+2}\right)$ Myr \\
Radial period & 239 & $237_{-5}^{+7}\left(_{-50}^{+2}\right)$ Myr \\
Vertical period & 346 & $342_{-7}^{+11}\left(_{-75}^{+2}\right)$ Myr \\
Mass of the progenitor & & $\sim2\times10^{4}$ $M_\sun$ \\
\enddata
\tablenotetext{a}{Maximum a posterior value, where available.}
\tablenotetext{b}{The median and the central 68\% confidence intervals are
measured from marginal posterior distributions. The intervals are calculated
from the difference of the 16th and 50th, and 84th and 50th percentile.}
\tablenotetext{c}{Recall that $E(B-V)_{off}$ is the offset with respect to
the reddening provided by the \citet{SFD98} dust map, and is {\em not} reddening
by itself.}
\tablenotetext{d}{The numbers in parenthesis illustrate the range of values
(with respect to the median) obtained when varying the distance of the Sun from
the Galactic center and the circular velocity at solar radius (see
Section~\ref{orbit}).}
\end{deluxetable}

\subsection{Element abundances\label{abundances}}

The preliminary element abundances of five red giant branch (RGB) stars in the
Ophiuchus stream (that is, the $v_{los}\sim290$ km s$^{-1}$ group), and observed
by Hectochelle, are listed in Table~\ref{table_abundances}. The uncertainties of
the determined parameters are listed in the notes of
Table~\ref{table_abundances}.

\begin{deluxetable}{lrrrrrr}
\tabletypesize{\scriptsize}
\setlength{\tabcolsep}{0.02in}
\tablecolumns{7}
\tablewidth{0pc}
\tablecaption{Element abundances of Ophiuchus stream stars\label{table_abundances}}
\tablehead{
\colhead{Name} & \colhead{$T_{eff}$} & \colhead{$\log g$} &
\colhead{${\rm [Fe/H]}$} & \colhead{${\rm [Mg/Fe]}$} &
\colhead{${\rm [Ca/Fe]}$} & \colhead{${\rm [Ti/Fe]}$}  \\
\colhead{ } & \colhead{(K)} & \colhead{(dex)} &
\colhead{(dex)} & \colhead{(dex)} & \colhead{(dex)} &
\colhead{(dex)}
}
\startdata
rgb1 & 5680 & 3.2 & -2.00 & -0.12 & 0.63 & 0.69 \\
rgb2 & 5430 & 3.3 & -1.95 &  0.06 & 0.62 & 0.47 \\
rgb3 & 5700 & 3.1 & -1.95 &  0.23 & 0.27 & 0.46 \\
rgb4 & 5400 & 2.8 & -1.90 &  0.24 & 0.57 & 0.42 \\
rgb5 & 5720 & 2.9 & -1.90 & -0.14 & 0.83 & -- 
\enddata
\tablecomments{The uncertainty in $T_{eff}$ is $<200$ K, $<0.45$ dex for
$\log g$, $\lesssim0.2$ dex for ${\rm [Fe/H]}$, and $\sim0.35$ dex for
abundances of $\alpha-$elements.}
\end{deluxetable}

We find the stars in the Ophiuchus stream to be poor in Fe
(${\mathrm [Fe/H]\sim-2.0}$ dex) and enhanced in $\alpha$-elements
(${\mathrm [\alpha/Fe]=0.4\pm0.1}$ dex). Their ${\mathrm [Fe/H]}$ are consistent
within 0.05 dex (rms scatter), despite fairly large estimated uncertainties in
individual measurements ($\lesssim0.2$ dex). The small scatter in
${\mathrm [Fe/H]}$ suggests that these stars come from the same single stellar
population.

Based on their position (within $5\arcmin$ of the Ophiuchus stream, as traced
by \citealt{ber14b}), kinematic and chemical properties, we conclude that all
stars listed in Table~\ref{table1} are high-probability members of the Ophiuchus
stream.

\subsection{Color-magnitude diagram}\label{CMD}

The sample of Ophiuchus stream members, which we have identified above using
velocities and metallicities, now gives us an opportunity to further constrain
the distance and the CMD of the stream.

\subsubsection{Model}

To model the CMD of the stream, we use a probabilistic approach analogous to the
one described in Section~\ref{vlos}. In our data set, $\mathcal{D}$, each data
point ${\bf d_k}$ is now defined by its galactic longitude and latitude, and by
its PS1 $grizy_{P1}$ magnitudes, ${\bf d_k}=\{\ell_k,b_k,g_k,r_k,i_k,z_k,y_k\}$.

Our data set contains only the Ophiuchus stream stars that were identified based
on spectroscopic data (i.e., velocity and metallicity). Thus, the set is
uncontaminated but very sparse and has a complicated spatial selection function.
Because of this, and because we are primarily interested in constraining the
distance of the stream, we focus on finding the isochrone(s) that match the
confirmed members in the color-magnitude space, and do not to model the
projected shape of the stream on the sky (for now, but see
Section~\ref{proper_motion_extent}).

To model the stream in color-magnitude space, we use a grid of theoretical
PARSEC isochrones\footnote{\url{http://stev.oapd.inaf.it/cmd}} (release v1.2S;
\citealt{bre12}; \citealt{che14}). Each isochrone $\mathcal{I}$ provides PS1
magnitudes
$m^\prime=g_{P1}^\prime,r_{P1}^\prime,i_{P1}^\prime,z_{P1}^\prime,y_{P1}^\prime$
for a star of initial mass $M_{init}$ in a single stellar population of age $t$,
metal content $Z$, and parametrized for the mass-loss on the RGB using the
variable values of the Reimers law parameter $\eta$ \citep{rei75, rei77}. At the
reference galactic longitude $\ell_0 = 5\arcdeg$, the distance modulus of the
stellar population is defined with parameter $\overline{DM}$, and a gradient in
distance modulus with galactic longitude is modeled with parameter
$\frac{dDM}{d\ell}$. We model the extinction in a PS1 band by adding the
\begin{equation}
    C_{ext}\cdot\left[E_{SFD}(B-V|\ell_k, b_k) + E(B-V)_{off}\right]
\end{equation}
term to isochrone magnitudes, where $E_{SFD}(B-V|\ell_k, b_k)$ is the reddening
at position $(\ell_k, b_k)$ in the \citet{SFD98} dust map, and
$E(B-V)_{off}$ accounts for a possible zero-point offset. The extinction
coefficients $C_{ext}=\{3.172, 2.271, 1.682, 1.322,1.087\}$ for PS1
$\{g,r,i,z,y\}$ bands are taken from Table 6 of \citet{sf11}. To account for the
fact that stellar evolution models are not perfect, we introduce five
$\sigma_{m,iso}$ parameters, where $m=g,r,i,z,y$, that model the uncertainty in
each PS1 $grizy$ magnitude provided by PARSEC isochrones.

Given the above model of the stream, the likelihood of a star $k$ with this
model is defined by the comparison of the spectral energy distribution
\{$g,r,i,z,y\}_{P1}$ with the prediction
\{$g^\prime,r^\prime,i^\prime,z^\prime,y^\prime\}_{P1}$ of an isochrone
$\mathcal{I}_k$ for a star with the initial mass $M_{init}$. Therefore,
\begin{equation}
    p({\bf d_k} | \mathcal{I}_k) = \int \left[\prod_{m=g,r,i,z,y} \mathcal{N}(m_k | m^\prime(M_{init}), \sigma^\prime_{m_k})\right]dM_{init}, \label{likelihood}
\end{equation}
where
\begin{equation}
    \mathcal{I}_k = \mathcal{I}(\ell_k, b_k, t, Z, \eta, E(B-V)_{off}, \overline{DM}, \frac{dDM}{d\ell}, {\mathbf \sigma_{iso}})
\end{equation}
is the isochrone at the galactic position of star $k$,
${\mathbf \sigma_{iso}}=\{\sigma_{g,iso}, \sigma_{r,iso}, \sigma_{i,iso}, \sigma_{z,iso}, \sigma_{y,iso}\}$, $\mathcal{N}(x|\mu,\sigma)$
is a normal distribution, and
\begin{equation}
    \sigma^\prime_{m_k}=\sqrt{\sigma_{m_k}^2 + \sigma_{m,iso}^2 + \left[0.1C_{ext}E_{SFD}(B-V|\ell_k, b_k)\right]^2}
\end{equation}
is the sum of the uncertainty in the isochrone magnitude ($\sigma_{m,iso}$),
observed magnitude of data point $k$ ($\sigma_{m_k}$), and extinction (10\%
fractional uncertainty in $E_{SFD}(B-V)$; \citealt{SFD98}). The likelihood of
all data points can then be calculated by combining
Equations~\ref{total_likelihood} and~\ref{likelihood}.

\subsubsection{Priors\label{priors_section} on the CMD model}

Before we can calculate the probability of a model, we need to define the prior
probabilities of model parameters. Below, we list our priors and describe the
justification for each one. A summary of priors is given in Table~\ref{priors}.

Based on spectroscopic data, the Ophiuchus stream is metal-poor
(${\rm [Fe/H]=-1.95\pm0.05}$ dex) and $\alpha-$enhanced
(${\rm [\alpha/Fe]=0.4\pm0.1}$ dex). As shown by \citet{scs93}, the
$\alpha-$enhanced stellar population models are equivalent to scaled-solar ones 
with the same global metal content ${\rm [M/H]}$, where ${\rm [M/H]}$ for
$\alpha-$enhanced models can be calculated using their Equation 3
\begin{equation}
    [M/H]\approxeq [Fe/H] + \log_{10}(0.638\times10^{[\alpha/Fe]} + 0.362).
\end{equation}
For the element abundance of the Ophiuchus stream, we estimate
${\rm [M/H]}\sim -1.7\pm0.2$ dex. This means that we should adopt the normal
distribution $\mathcal{N}(\log_{10}(Z/Z_\sun)| -1.7, 0.2)$ as the prior
probability of metallicity $Z$ (where $Z_\sun=0.0152$ is the solar metal content
used by this perticular set of PARSEC isochrones). However, in the context of
cross-validating our analysis, we decided to replace the above metallicity prior
in favor of a (less informative) prior that is uniform in the
$0.0001 < Z < 0.0004$ range. Even though a less informative prior was adopted,
at the end of Section~\ref{CMD_posterior} we find a very impressive consistency
between the posterior distribution of metallicity $Z$ (obtained using CMD
fitting) and the spectroscopic estimate of $Z$ (see bottom panel of
Figure~\ref{DM_Z_posteriors}). In the end, it is important to note that our
results would not have changed significantly if we used the more informative
prior for metallicity content $Z$.

The presence of BHB stars, the ${\rm [Fe/H]}$ and the $\alpha-$enhancement of
the stream point to an old stellar population. Thus, for age we adopt a uniform
prior in the $8 < t/{\rm Gyr} < 13.5$ range.

Metal-poor and old populations have $\eta\sim0.4$ \citep{rpf88}. Therefore, for
the mass-loss parameter $\eta$ we adopt a uniform prior in the $0.2<\eta<0.5$
range. For the uncertainty in isochrone magnitudes, we adopt a prior that is
uniform in the $0 \leq \sigma_{m,iso} < 0.1$ mag range, where $m=g,r,i,z,y$.

According to \citet{ber14b}, the Ophiuchus stream is located about $9\pm1$ kpc
from the Sun. Thus, for $\overline{DM}$ we adopted a uniform prior in the
$14.2 < \overline{DM} < 15.2$ mag range (corresponding to the 7-11 kpc range).
For the gradient in distance modulus, a uniform prior in the
$\arrowvert\frac{dDM}{d\ell}\arrowvert < 0.5$ mag deg$^{-1}$ range is adopted.
Since the reddening in the region of interest is greater than 0.1 mag, we assume
that the possible systematic offset in $E(B-V)$ values provided by the
\citet{SFD98} dust map is less than 0.1 mag (i.e., $|E(B-V)_{off}|<0.1$
mag).

\begin{deluxetable}{lll}
\tabletypesize{\scriptsize}
\setlength{\tabcolsep}{0.02in}
\tablecolumns{3}
\tablewidth{0pc}
\tablecaption{Prior probabilities of CMD parameters\label{priors}}
\tablehead{
    \colhead{Parameter} & \colhead{Prior type} & \colhead{Range}
}
\startdata
Age $t$ & uniform & 8 to 13.5 Gyr\\
Mass-loss parameter $\eta$ & uniform & 0.2 to 0.5 \\
Metallicity $Z$ & uniform & 0.0001 to 0.0004 \\
$E(B-V)_{off}$ & uniform & -0.1 to 0.1 mag \\
$\overline{DM}$ & uniform & 14.2 to 15.2 mag\\
$\frac{dDM}{d\ell}$ & uniform & -0.5 to 0.5 mag deg$^{-1}$ \\
$\sigma_{m,iso}$ ($m=g,r,i,z,y$)& uniform & 0 to 0.1 mag
\enddata
\end{deluxetable}

\subsubsection{Posterior distributions of CMD parameters\label{CMD_posterior}}

To efficiently explore the parameter space, we again use the \texttt{emcee}
package. We use 1000 walkers and obtain convergence after a short burn-in phase
of 100 steps per walker. The chains are then restarted around the best-fit value
and evolved for another 4000 steps. The maximum a posterior values, the median
and the central 68\% confidence intervals of model parameters are listed in
Table~\ref{stream_parameters}.

\begin{figure}
\plotone{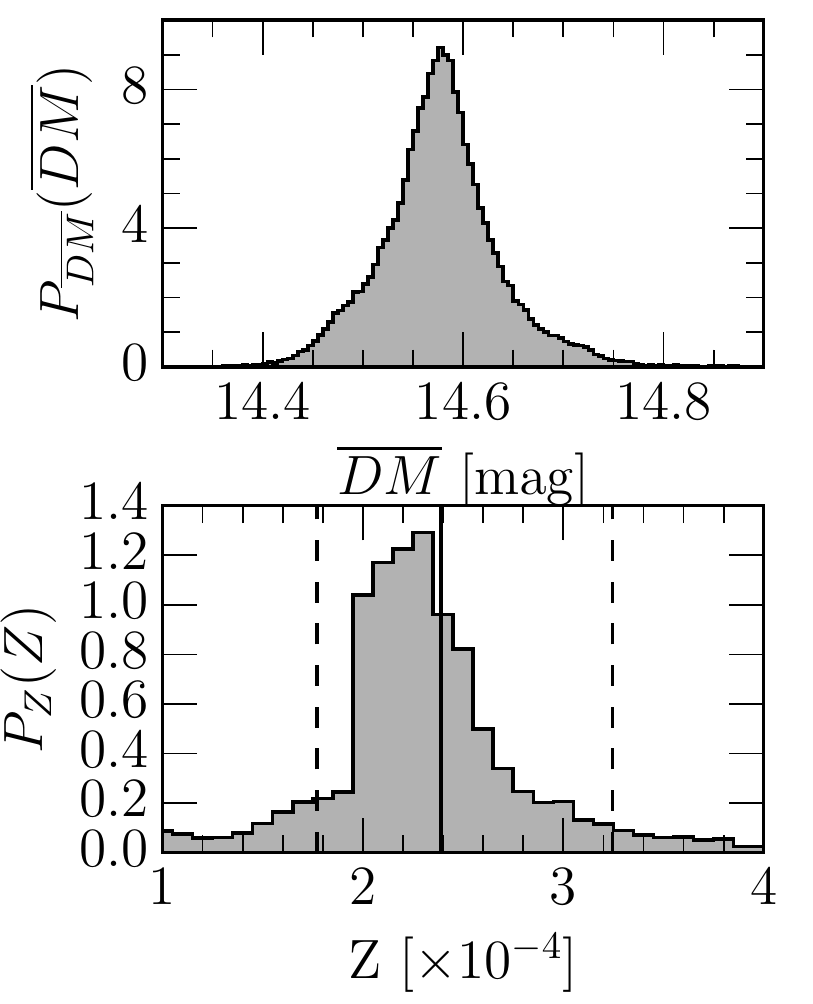}
\caption{
Marginal posterior distributions of distance modulus at $\ell_0=5\arcdeg$
({\em top})
and metallicity $Z$ ({\em bottom}). In the bottom panel, the solid vertical line
shows the mean metallicity $Z$ measured from spectroscopy
(${\rm [Fe/H]=-1.95\pm0.05}$ dex, ${\rm [\alpha/Fe]} = 0.4\pm0.1$ dex), while
the dashed lines show the uncertainty in the spectroscopic estimate of $Z$.
\label{DM_Z_posteriors}}
\end{figure}

\begin{figure}
\plotone{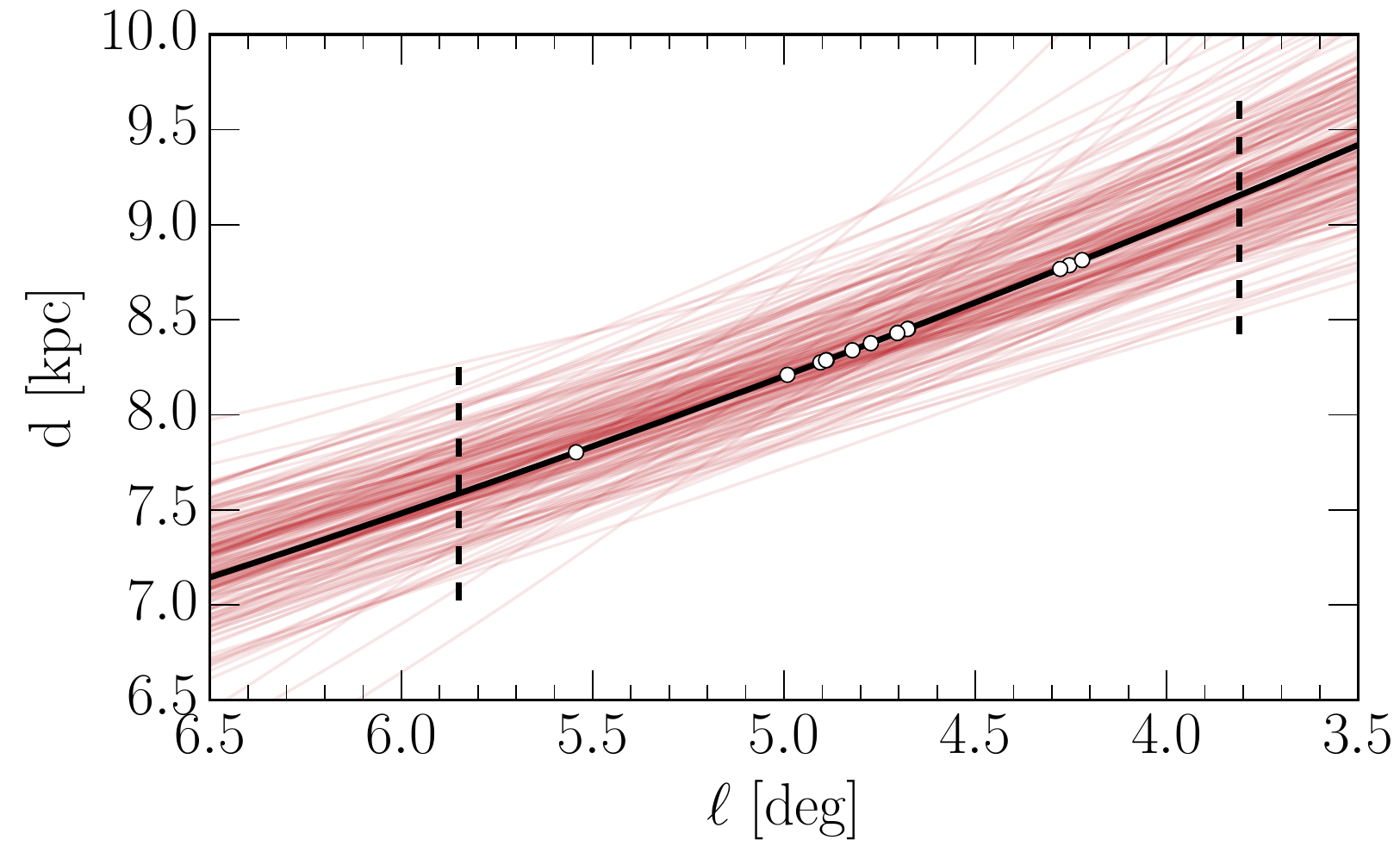}
\caption{
Heliocentric distance of the Ophiuchus stream as a function of galactic
longitude $\ell$. The thick solid line shows the most probable model,
$DM(\ell)=-0.20(\ell-5)+14.57$ mag. To illustrate the uncertainty in the most
probable model, the thin semi-transparent red lines show 200 models sampled from
the posterior distribution. The vertical dashed lines show the likely extent of
the stream (see Section~\ref{proper_motion_extent}). The white circles plotted
on top of the solid line show the positions of 14 confirmed stream members,
where their distance modulus was calculated using the most probable model of
$DM(\ell)$.
\label{gl_vs_distance}}
\end{figure}

We find the stream to be $\sim12$ Gyr old and to have a distance modulus of
$14.58\pm0.05$ mag (i.e., a distance of 8.2 kpc) at the reference galactic
longitude $l_0=5\arcdeg$ (top panel of Figure~\ref{DM_Z_posteriors}). Most
importantly, we detect a gradient of $-0.20\pm0.03$ mag deg$^{-1}$ in distance
modulus. This gradient is inconsistent with zero (i.e., with the no gradient
hypothesis) at a $7\sigma$ level, and confirms the suggestion by \citet{ber14b}
that the eastern part of the stream is closer to the Sun
(Figure~\ref{gl_vs_distance}). The gradient in distance modulus is mostly
constrained by BHB and MSTO/SGB stars. When these stars are not used to
constrain the CMD of the stream, the marginal posterior distribution of the
gradient in distance modulus becomes multimodal and poorly constrained.
    
To verify whether the observed gradient in distance modulus is real, we compared
derredened colors and magnitudes of BHB stars in the Ophiuchus stream. Stars
bhb1 and bhb3 have identical dereddened $g_{P1}-i_{P1}$ color
($g_{P1}-i_{P1}=-0.43$ mag), and thus should have identical absolute
magnitudes\footnote{Assuming they are members of a single stellar population,
which seems to be the case based on a lack of intrinsic scatter in
${\rm [Fe/H]}$ of RGB stars (see Section~\ref{abundances}).}. However, their
dereddened $i_{P1}$-band magnitudes differ by 0.1 mag, and the brighter star in 
the pair is located 0.5 deg eastward (in the galactic longitude direction), in
agreement with the observed distance modulus gradient of -0.2 mag deg$^2$. The
stars bhb6 and bhb7 also show similar behavior.

\begin{figure}
\plotone{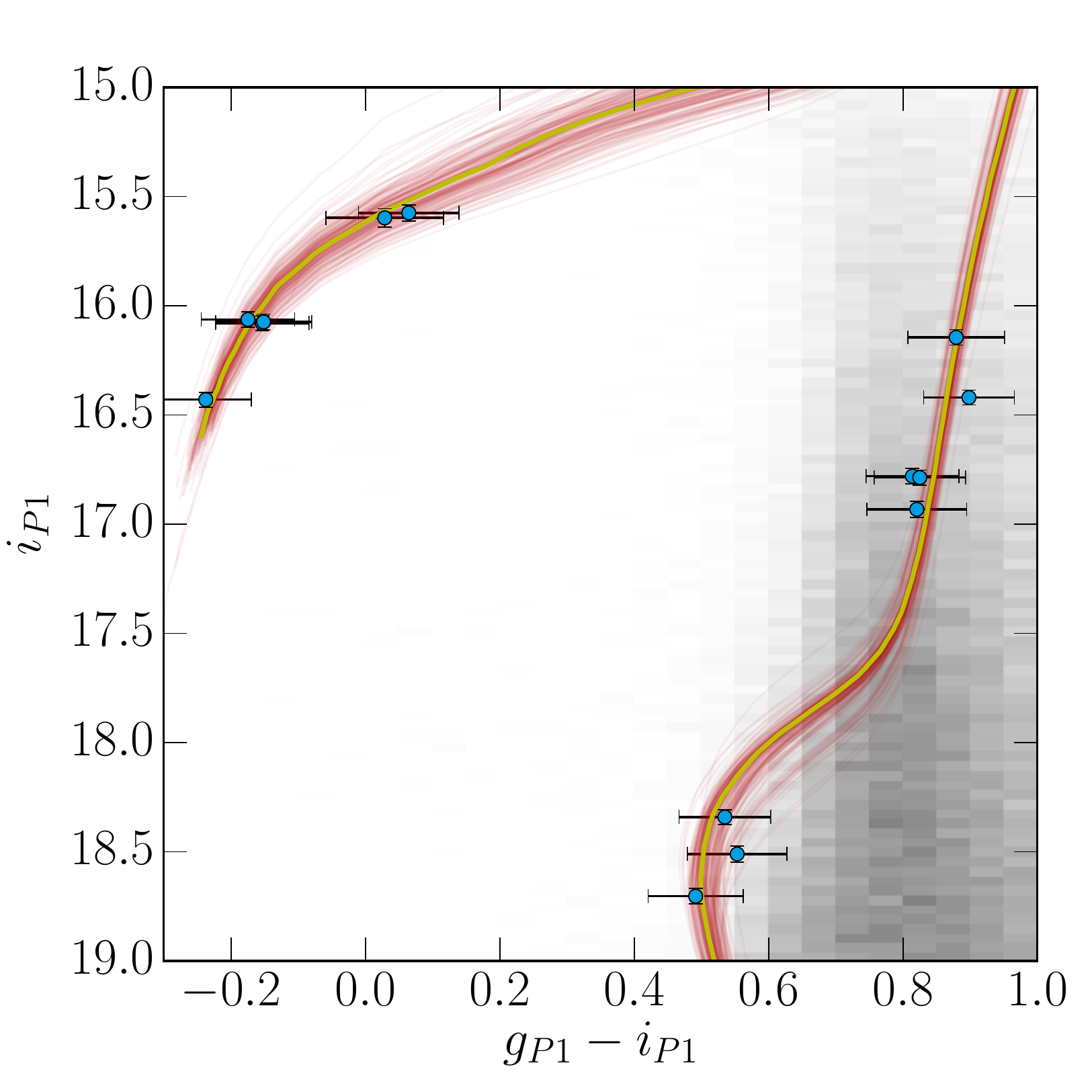}
\caption{
The $g_{P1}-i_{P1}$ vs.~$i_{P1}$ color-magnitude diagram showing the most
probable isochrone (yellow thick line) and 200 isochrones randomly sampled from 
the stream's full posterior distribution (semi-transparent dark red thin lines).
The isochrones have been shifted to match the distance of the stream at
$\ell_0=5\arcdeg$, and have been reddened assuming $E(B-V)=0.19$ mag (median
$E_{SFD}(B-V)$ at the position of the stream). The grayscale pixels show the
density distribution of field stars in this diagram (i.e., their probability
density function). For illustration only, the magnitudes of observed stars have
been corrected for extinction using the \citet{SFD98} dust map (to correct for
gradients in reddening), and then again extincted assuming $E(B-V)=0.19$ mag. In
addition, the $i_{PS1}$-band magnitudes of observed stars have been corrected
for the gradient in distance modulus by adding
$\frac{dDM}{d\ell}(\ell-\ell_0)=-0.2(\ell-5)$ mag. Note that the uncertainties
in color and magnitude also include the uncertainty in extinction.
\label{CMD_models}}
\end{figure}

In Figure~\ref{CMD_models}, we compare CMDs of the Ophiuchus stream and field
stars. The CMD of field stars (grayscale pixels) was obtained by binning the
$g_{P1}-i_{P1}$ colors and $i_{P1}$-band magnitudes of stars located more than
$18\arcmin$ from the equator\footnote{See Section 3 of \citealt{ber14b} for its
definition.} of the Ophiuchus stream.

In Section~\ref{priors_section}, we adopted a uniform prior for the metallicity
content $Z$ in order to test the predictive power our dataset. As shown in the
bottom panel of Figure~\ref{DM_Z_posteriors}, the peak of the marginal posterior
distribution of $Z$ is consistent with the mean value of $Z$ estimated from
spectroscopic data (solid vertical line), and the distribution is even narrower 
than the distribution of $Z$ estimated from spectroscopy (dashed vertical
lines). This result demonstrates the predictive power of our data. It shows how 
a combination of good coverage of the CMD, PS1 photometry, and detailed modeling
can provide an accurate and precise estimate of the metallicity of single
stellar populations.

\subsection{Modeling the proper motion and the extent of the stream\label{proper_motion_extent}}

The longitude-dependent CMD model we have built in Section~\ref{CMD}, and the
luminosity functions associated with the model, allow us to assign a likelihood
that a star is a member of the Ophiuchus stream, based on the star's galactic
longitude $\ell$, $g_{P1}-i_{P1}$ color, and $i_{P1}$-band magnitude. The
distribution of field stars in the $g_{P1}-i_{P1}$ vs.~$i_{P1}$ CMD (grayscale
pixels in Figure~\ref{CMD_models}), on the other hand, enables us to estimate
the likelihood that a star is associated with the field. As we show in this
section, these two probability density functions (PDFs), when combined with
positional and proper motion data, can be used to simultaneously trace the
extent of the Ophiuchus stream and determine its proper motions across the sky.

In principle, we could measure the proper motion of the Ophiuchus stream using
the proper motion of its confirmed members. However, since our sample of
confirmed members contains only 14 stars, there is a possibility that one or two
stars with incorrectly measured proper motions may bias the results. As an
example, stream member ``rgb4'' is clearly an outlier in proper motion as it has
$\mu_\ell\sim-24$ mas yr$^{-1}$, while the remaining members have
$\mu_\ell\sim-6$ mas yr$^{-1}$. A visual inspection of digitized photographic
plates has revealed that ``rgb4'' is blended with a neighbor of similar
brightness, which affects the measured position of the star and its proper
motion.

Fortunately, we do not need to rely only on confirmed members and can use a much
larger sample of stars in the vicinity of the Ophiuchus stream to constrain its
proper motion and extent. As we detail below, we use a probabilistic approach
(see Sections~\ref{vlos} and~\ref{CMD}) and model the distribution of stars
simultaneously in coordinate, proper motion, and color-magnitude space as a
mixture of stream and field (i.e., non-stream) stars. Even though we do not
{\em a priori} know which star is a true member of the stream, we assume that as
an ensemble, the stream stars have certain characteristics which make them
distinguishable from field stars (e.g., common proper motion, distance, position
on the sky and in the CMD), and that the {\em scatter} in these characteristics
is sufficiently small to overcome the fact that there are a lot more field than
stream stars. The narrow width of the stream in color-magnitude
(Figure~\ref{CMD_models}) and coordinate space (Figure~1 of \citealt{ber14b})
support this assumption. After all, if the stream did not have these
characteristics, it likely would not have been detected by \citet{ber14b} in the
first place.

Even though the probabilistic approach we describe below uses all of the stars
in the vicinity of the Ophiuchus stream to constrain the its extent and proper
motion, we expect that most of the signal will come from main sequence turn-off
(MSTO) stars associated with the stream. As shown in Figure~\ref{CMD_models},
the stream's MSTO is bluer than the field population, which means that stars in
this region of the CMD are much more likely to be associated with the stream
than with the field population. Thus, the statistical weight of such stars will
be greater than, for example, the weight of stream's RGB stars, which occupy the
region of the CMD that is heavily dominated by field stars.

Assuming the stream extends between galactic longitudes $\ell_{min}$ and
$\ell_{max}$, the likelihood that a star with galactic longitude $\ell_k$,
latitude $b_k$, proper motions in galactic coordinates of $\mu_{\ell,k}$ and
$\mu_{b,k}$, color $(g-i)_k$ and magnitude $i_k$ is drawn from the mixture
model, is equal to
\begin{equation}
p({\bf d_k} | \theta, \ell_k) = fp_{str}({\bf d_k} | \theta_{str}, \ell_k) + (1-f)p_{fld}({\bf d_k} | \theta_{fld}, \ell_k),\label{mix_model}
\end{equation}
where ${\bf d_k} \equiv \{\ell_k, b_k, \mu_{\ell,k}, \mu_{b,k}, (g-i)_k, i_k \}$
contains measurements for data point (star) $k$, and
$\theta\equiv\{\theta_{str}, \theta_{fld}\}$ contains parameters that model
the distribution of stream and field stars, respectively. The parameter $f$
specifies the fraction of stars in the stream (out of all stars between
$\ell_{min}$ and $\ell_{max}$) and is $0\le f\le1$ for all $\ell_k$ where
$\ell_{min} < \ell_k < \ell_{max}$, otherwise, it is $f=0$.

The likelihood $p_{str}({\bf d_k} | \theta_{str}, \ell_k)$ is a product of
spatial likelihood $p_{str}^{sp}$, proper motion likelihood $p_{str}^{pm}$, and
the color-magnitude likelihood $p_{str}^{cm}$
\begin{equation}
\begin{split}
p_{str}({\bf d_k} |\theta_{str}, \ell_k)=p_{str}^{sp}(b_k|\theta^{sp}_{str}, \ell_k) \\
\times p_{str}^{pm}(\mu_{\ell,k},\mu_{b,k}|\theta^{pm}_{str}, \ell_k) \\
\times p_{str}^{cm}((g-i)_k, i_k| \theta^{cm}_{str}, \ell_k, b_k, E_{SFD}(B-V|\ell_k, b_k)).
\end{split}
\end{equation}
The likelihood for field stars, $p_{fld}({\bf d_k} | \theta_{fld}, \ell_k)$, has
the same decomposition.

In galactic coordinates, the distribution of stream stars in the latitude
direction is modeled with a Gaussian of width $\sigma_b$, where the latitude
position of the Gaussian changes as a quadratic function of the galactic
longitude
\begin{equation}
p_{str}^{sp}(b_k | \theta^{sp}_{str}, \ell_k) = \mathcal{N}(b_k | \nu(\ell_k), \sigma_b)\label{stream_spatial_likelihood},
\end{equation}
where $\nu(\ell_k | A, B, C)=A + B(\ell_k-\ell_0) + C(\ell_k-\ell_0)^2$ is the
galactic latitude of the equator of the stream and $\ell_0=5\arcdeg$.

The spatial distribution of field stars is modeled in a similar fashion, except
the Gaussian has a width $\sigma^\prime_b$, and its latitude position is a
linear function of the galactic longitude
\begin{equation}
p_{fld}^{sp}(b_k | \theta^{sp}_{fld}, \ell_k) = \mathcal{N}(b_k | \nu^\prime(\ell_k), \sigma^\prime_b),
\end{equation}
where $\nu^\prime(\ell_k|A^\prime,B^\prime)=A^\prime + B^\prime(\ell_k-\ell_0)$.
We use the above model for $p_{fld}^{sp}$ because it is easy to implement, and
because for large ratios of $\sigma^\prime_b/\sigma_b$ the Gaussian that models
the spatial distribution of field stars approximates to a plane with respect to
the much narrower Gaussian that describes the spatial distribution of stream
stars.

At the reference galactic longitude $\ell=5\arcdeg$, the stream is assumed to
have proper motion $\overline{\mu_\ell}$ and $\overline{\mu_b}$, with possible
gradients in proper motion of $\frac{d\mu_\ell}{d\ell}$ and
$\frac{d\mu_b}{d\ell}$ (i.e., gradients as a function of galactic longitude).
The proper motion likelihood of stream stars is then
\begin{equation}
\begin{split}
p_{str}^{pm}(\mu_{\ell,k},\mu_{b,k}|\theta^{pm}_{str}, \ell_k)= \\
\mathcal{N}(\mu_{\ell,k}| \mu_\ell(\ell_k), \sigma^\prime_k) \\
\times\mathcal{N}(\mu_{b,k}| \mu_b(\ell_k), \sigma^\prime_k),
\end{split}
\end{equation}
where $\mu_\ell(\ell_k)=\frac{d\mu_\ell}{d\ell}\left(\ell_k-\ell_0\right) + \overline{\mu_\ell}$ and
$\mu_b(\ell_k)=\frac{d\mu_b}{d\ell}\left(\ell_k-\ell_0\right)+\overline{\mu_b}$
are the predicted proper motions of the stream at galactic longitude $\ell_k$,
and $\sigma^\prime_k = \sqrt{\sigma_{pm}^2 + \sigma^2_{\mu,k}}$ is the quadratic
sum of the intrinsic proper motion dispersion and the uncertainty in the
corresponding proper motion of data point $k$. The purpose of parameter
$\sigma_{pm}$ is to account for any additional scatter in proper motions (e.g., 
due to unaccounted errors). The proper motion likelihood of field stars has the
same form (but different parameters) as the proper motion likelihood of stream
stars.

The likelihood that a star is drawn from the stream's CMD is defined as
\begin{equation}
\begin{split}
p_{str}^{cm}((g-i)_k, i_k| \theta^{cm}_{str}, \ell_k, b_k, E_{SFD}(B-V|\ell_k, b_k))= \\ \zeta \int\int \mathcal{N}((g-i)^\prime_k| g-i, \sigma_{(g-i)_k}) \\
\times \mathcal{N}(i^\prime_k | i, \sigma_{i_k})p(g-i, i | str)d(g-i)di,\label{CM_likelihood}
\end{split}
\end{equation}
where $\sigma_{(g-i)_k}$ and $\sigma_{i_k}$ are the uncertainty in color and
magnitude of data point $k$, and $\zeta$ is a normalization constant calculated 
such that the integral of Equation~\ref{CM_likelihood} over the considered
region of CM space is unity. To account for the extinction and the gradient in
distance, we use color $(g-i)^\prime_k=(g-i)_k-1.49E_{SFD}(B-V|\ell_k, b_k)$ and
magnitude
$i^\prime_k=i_k - 1.682E_{SFD}(B-V|\ell_k, b_k) - \frac{dDM}{d\ell}(\ell_k-5)$,
where $\frac{dDM}{d\ell}=-0.20$ mag deg$^{-1}$ is the most probable gradient in
distance modulus (see Table~\ref{stream_parameters}).

In Equation~\ref{CM_likelihood}, $p(g-i, i | \theta^{cm}_{str})$ is the
PDF of the Ophiuchus stream in the $g_{P1}-i_{P1}$ vs.~$i_{P1}$ color-magnitude
space at galactic longitude $\ell_0=5\arcdeg$. This PDF was constructed by
sampling isochrones from the stream's CMD model (Section~\ref{CMD_posterior}),
multiplying them with their luminosity functions, and then summing them up in a 
binned $g_{P1}-i_{P1}$ vs.~$i_{P1}$ CMD.

The likelihood that a star is drawn from the field CMD is calculated as
\begin{equation}
\begin{split}
p_{fld}^{cm}((g-i)_k, i_k| \theta^{cm}_{bkg}, \ell_k, b_k)= \\ \zeta \int\int \mathcal{N}((g-i)_k| g-i, \sigma_{(g-i)_k}) \\
\times \mathcal{N}(i_k | i, \sigma_{i_k})p(g-i, i | \theta^{cm}_{bkg}, \ell_k, b_k)d(g-i)di.\label{CM_likelihood_fld}
\end{split}
\end{equation}
In Equation~\ref{CM_likelihood_fld}, the observed color and magnitude are
{\em not} corrected for extinction or any gradients.

The PDF of field stars (i.e., the CMD),
$p(g-i, i|\theta^{cm}_{bkg},\ell_k, b_k)$ in Equation~\ref{CM_likelihood_fld},
depends on the galactic position. We construct
$p(g-i, i | \theta^{cm}_{bkg}, \ell_k, b_k)$ by dividing the Ophiuchus region
into $1\arcdeg\times1\arcdeg$ spatial pixels that overlap by $0.5\arcdeg$ in
galactic longitude and latitude directions. For each spatial pixel, we bin
$g_{P1}-i_{P1}$ colors and $i_{P1}$-band magnitudes of stars located more than
$18\arcmin$ from the equator\footnote{See Section 3 of \citealt{ber14b} for
its definition} of the stream, and normalize the resulting CMD to unity area. An
example CMD of field stars is shown in Figure~\ref{CMD_models} as grayscale
pixels.

In total, our model contains 20 parameters. For all of the parameters, we have
adopted uniform priors within reasonable bounds. The allowed ranges of model
parameters were determined by examining positions and proper motions of
confirmed members and other stars. In addition to adopted priors, we also
require that the parameters satisfy the following constraints:
\begin{enumerate}
\item the spatial width of the stream must be smaller or equal than the width of
the spatial distribution of field stars: $\sigma_b \leq \sigma^\prime_b$
\item the additional scatter in proper motion of stream stars must be smaller
than the scatter in proper motions of field stars:
$\sigma_{pm} \leq \sigma^\prime_{pm}$, and
\item the galactic latitudes of confirmed members ($b^{conf}_k$) must be within
$3\sigma_b$ of the equator of the stream:
$|\nu(l^{conf}_k) - b^{conf}_k| \leq 3\sigma_b$,
where $\nu(l^{conf}_k)$ is the galactic latitude of the stream at the position
of confirmed members, and $1 \leq k \leq 14$.
\end{enumerate}

As our data set, we use stars brighter than $i_{P1}=20$ mag with measured proper
motions, and located in a $4\times4$ deg$^2$ area centered on the Ophiuchus
stream. To explore the parameter space, we use 200 \texttt{emcee} walkers and
obtain convergence after a short burn-in phase of 100 steps. The chains are then
restarted around the best-fit value and evolved for another 2000 steps. The
maximum a posterior values, the median and the central 68\% confidence intervals
of model parameters are listed in Table~\ref{stream_parameters}.

\begin{figure}
\plotone{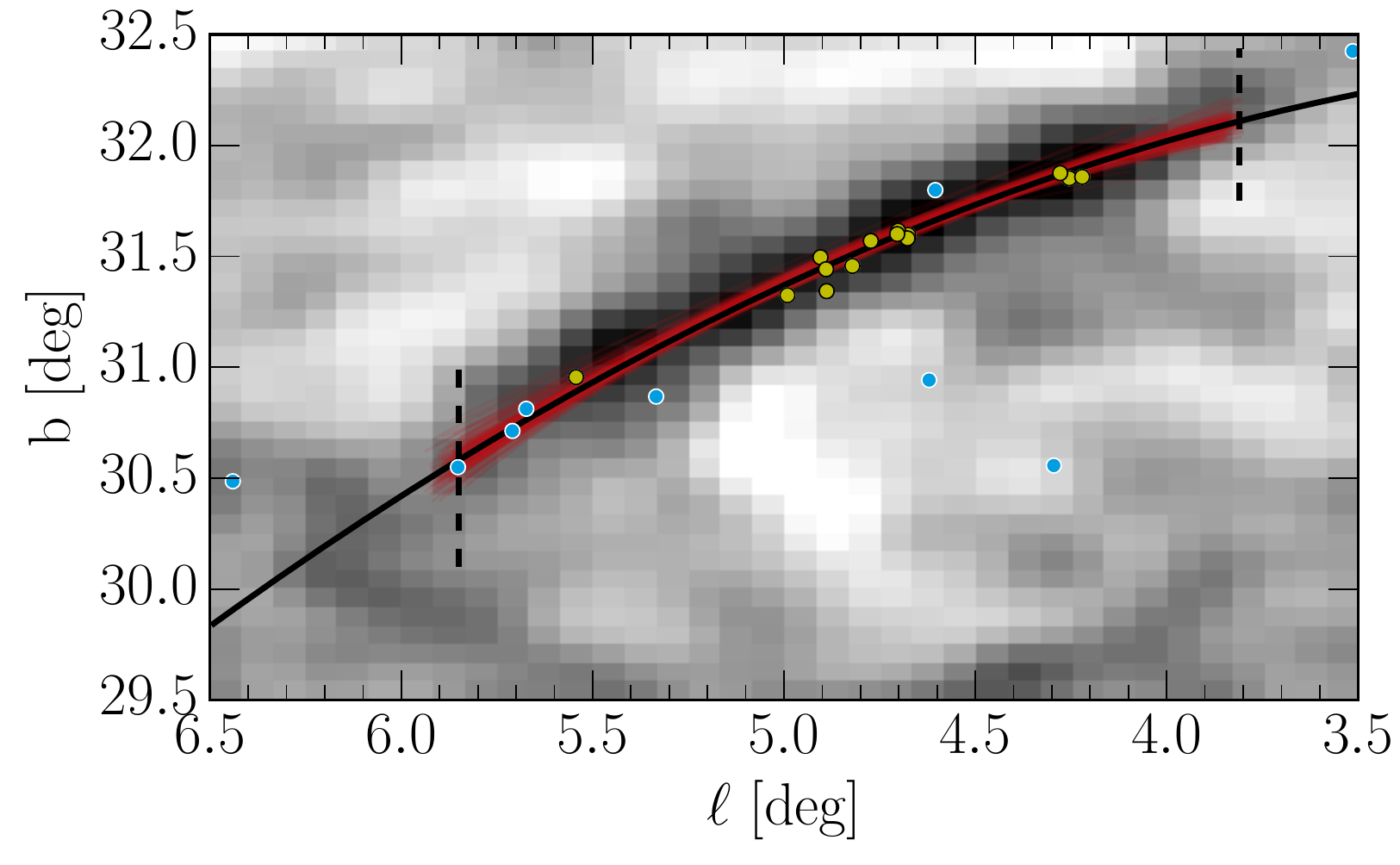}
\caption{
The extent of the Ophiuchus stream in galactic coordinates. The gray-scale map
shows the probability-weighted number density of the Ophiuchus stream, smoothed 
using a $6\arcmin$-wide Gaussian filter. The thick solid line shows the most
probable model ($b(\ell)=31.37-0.80(\ell-5)-0.15(\ell-5)^2$ deg) for the equator
of the stream. To illustrate the uncertainty in the most probable model, the
thin semi-transparent red lines show 200 models sampled from the posterior
distribution. The vertical dashed lines show the likely extent of the stream
(see Section~\ref{proper_motion_extent}). The yellow points show the positions
of confirmed members, the blue points show candidate BHB stars (probability of
being stream members $>80\%$), and the arrow indicates the direction of movement
of the stream.
\label{gl_vs_gb_track}}
\end{figure}

We find the stream to be confined between galactic longitudes of $3.81\arcdeg$
and $5.85\arcdeg$ (Figure~\ref{gl_vs_gb_track}). When combined with the distance
of the stream (Figure~\ref{gl_vs_distance}), this result implies that the
deprojected length of the stream is $1.6\pm0.3$ kpc. Thus, the stream is very
foreshortened in projection, by a ratio of $6:1$. The galactic latitude of the
equator of the stream is at $b_{stream}(\ell)=31.37-0.80(\ell-5)-0.15(\ell-5)^2$
deg, and the width of the stream is $\sigma_b\sim6$ arcmin (in the galactic
latitude direction). In direction perpendicular to the stream's equator, the
stream is $\sim3.5\arcmin$ wide, which is similar to width of $\sim3\arcmin$
measured by \citet{ber14b}.

Using the $g_{P1}-i_{P1}$ color and $i_{P1}$-band magnitude, and given the CMD
model of the stream (Section~\ref{CMD_posterior}), we can evaluate the
probability that a star is a member of the Ophiuchus stream. We have calculated
these probabilities for all of the stars in the vicinity of the Ophiuchus stream
and have created a probability-weighted number density map of the stream, shown
as grayscale pixels in Figure~\ref{gl_vs_gb_track}. An inspection of the number
density map did not reveal a significant overdensity of stars along the stream
that would indicate the presence of a progenitor.

\begin{figure}
\plotone{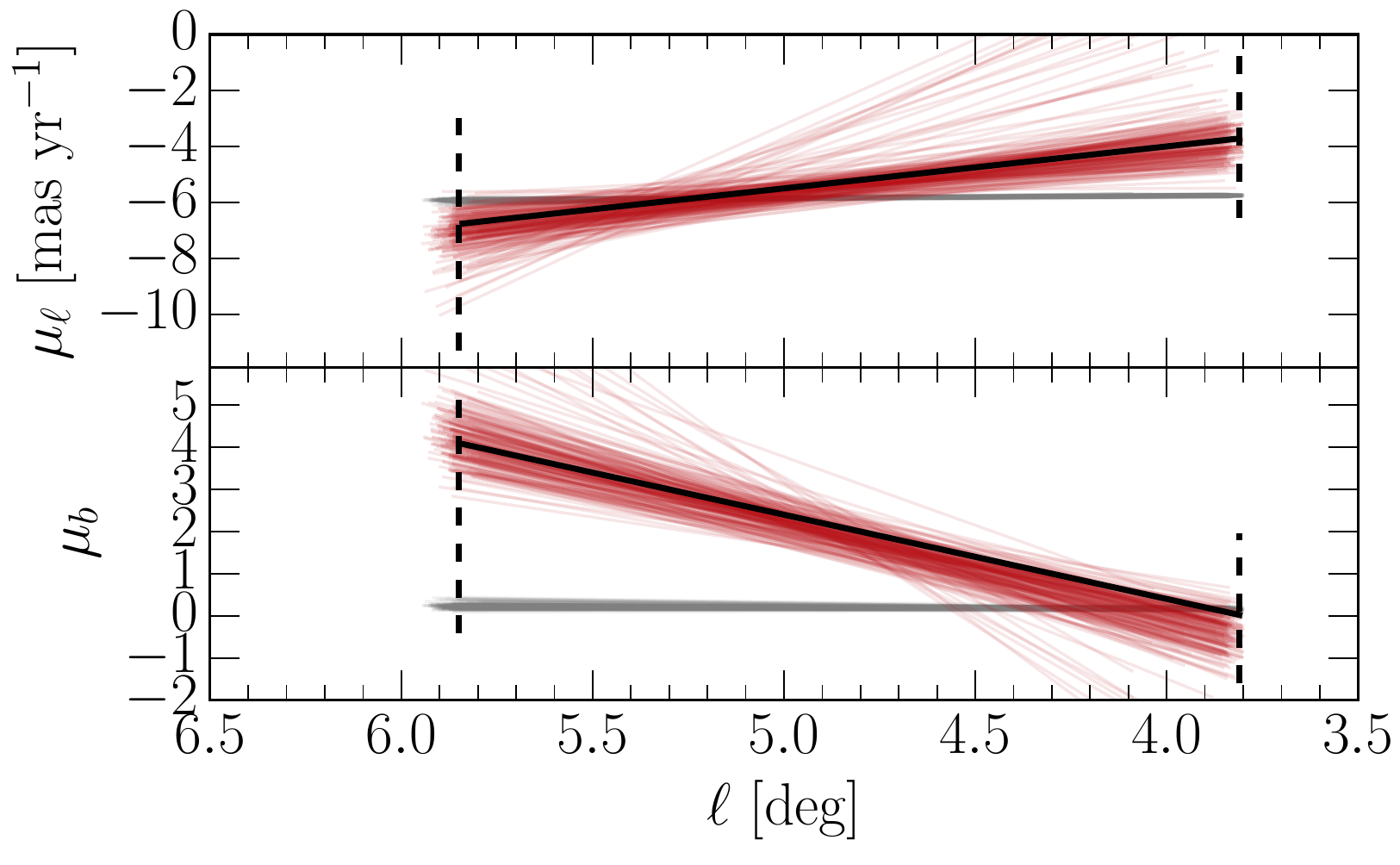}
\caption{
Proper motion of the Ophiuchus stream, inferred from the ensemble of likely
stream members. The panels show the proper motion in galactic longitude
({\em top}) and latitude directions ({\em bottom}), as a function of galactic
longitude $\ell$. The thick solid lines show the most probable models
($\mu_\ell(\ell)=-1.5(\ell-5)-5.5$ mas yr$^{-1}$, $\mu_b(\ell)=2.0(\ell-5)+2.4$
mas yr$^{-1}$). To illustrate the uncertainty in most probable models, the thin 
semi-transparent red lines show 200 models sampled from respective posterior
distributions. For comparison, the semi-transparent gray lines show the proper
motion of field stars. The vertical dashed lines show the likely extent of the
stream (see Section~\ref{proper_motion_extent}).
\label{gl_vs_pm}}
\end{figure}

Our data indicate that the proper motion of the stream changes as a function of
galactic longitude (Figure~\ref{gl_vs_pm}). The gradients in proper motion
are significant at $\gtrsim3\sigma$ level, and while their absolute values are
similar ($\sim2$ mas yr$^{-1}$ deg$^{-1}$), the gradients have opposite signs.
For comparison, the gradients in proper motions of field stars are 10 times
smaller, $\frac{d\mu^\prime_\ell}{d\ell}\sim0.1$ mas yr$^{-1}$ deg$^{-1}$ and
$\frac{d\mu^\prime_b}{d\ell}\sim0.2$ mas yr$^{-1}$ deg$^{-1}$. Overall,
the proper motions of field stars
($\mu^\prime_\ell\sim-6$ mas yr$^{-1}$ and $\mu^\prime_b\sim0.2$ mas yr$^{-1}$)
are consistent with apparent motions of a population at $\sim8$ kpc (due to the
motion of the Sun around the Galaxy). For comparison, the apparent motion of the
compact radio source Sgr A$^*$ at the Galactic center is
$\mu^{SgrA^*}_{\ell}=-6.38$ mas yr$^{-1}$ and $\mu^{SgrA^*}_{b}=-0.20$ mas
yr$^{-1}$ \citep{rb04}.

The stream parameters we have obtained so far can be used to place a lower limit
on the mass of the initial population of the Ophiuchus stream. The fraction of
stars $f$ in the stream between longitudes $\ell_{min}$ and $\ell_{max}$, can be
converted to the number of stars in the stream, $N_{stars}$. We find that there
are $N_{stars}=300\pm30$ stars brighter than $i_{P1}=20$ mag in the Ophiuchus
stream. If we adopt the luminosity function associated with the most probable
CMD model of the stream and assume \citet{kro98} initial mass function (not
corrected for binarity), this number of stars implies that the initial
population of the Ophiuchus stream had to have a mass of at least
$M_{init}=(7.0\pm0.7)\times10^3$ $M_{\sun}$.

\section{Orbit of the Ophiuchus stream}\label{orbit}

The data and models of the stream obtained in previous sections now enable us to
constrain the orbit of the Ophiuchus stream. For this purpose we use
\texttt{galpy}\footnote{\url{http://github.com/jobovy/galpy}}, a package for
galactic dynamics written in \texttt{Python} programming language \citep{bov15}.

To make the best use of the stream constraints and their covariances derived so
far, we sample the PDF of stream constraints with 200 stream samples. Each of
these samples consists of the line of sight velocity, distance modulus, galactic
position, and proper motion for 14 stars that uniformly sample the stream in
galactic longitude from $\ell_{min}$ to $\ell_{max}$, where these two values are
drawn for each of the 200 samples from the posterior distribution.

To emulate the width of the stream, we assign an uncertainty of $\sigma_b$ to
positions of data points. To all data points we assign a 3\% uncertainty in
distance, 2 km s$^{-1}$ uncertainty in velocity, and 2 mas yr$^{-1}$ of
uncertainty in proper motions. We have verified that our results do not change
significantly if these uncertainties are changed within reason. To convert the
observed values into 3D positions and velocities, \texttt{galpy} assumes the Sun
is located 8 kpc from the Galactic center ($R_0=8$ kpc), the circular velocity
at the solar radius is $v_{circ}(R_0)=220$ km s$^{-1}$, and Sun's motion in the
Galaxy is (-11.1, 244, 7.25) km s$^{-1}$ \citep{sbd10, bov12}.

We fit the orbit of each of the 200 stream samples. The orbits are integrated in
the default \texttt{galpy} potential, called \texttt{MWPotential2014} (Table 1
of \citealt{bov15}). This potential consists of a bulge modeled as a power-law
density profile that is exponentially cutoff with a power-law exponent of -1.8
and a cut-off radius of 1.9 kpc, a Miyamoto-Nagai disk, and a dark-matter NFW
halo. \texttt{MWPotential2014} is consistent with a large variety of dynamical
constraints on the potential of the Milky Way, ranging from the bulge to the
outer halo.

\begin{figure*}
\plottwo{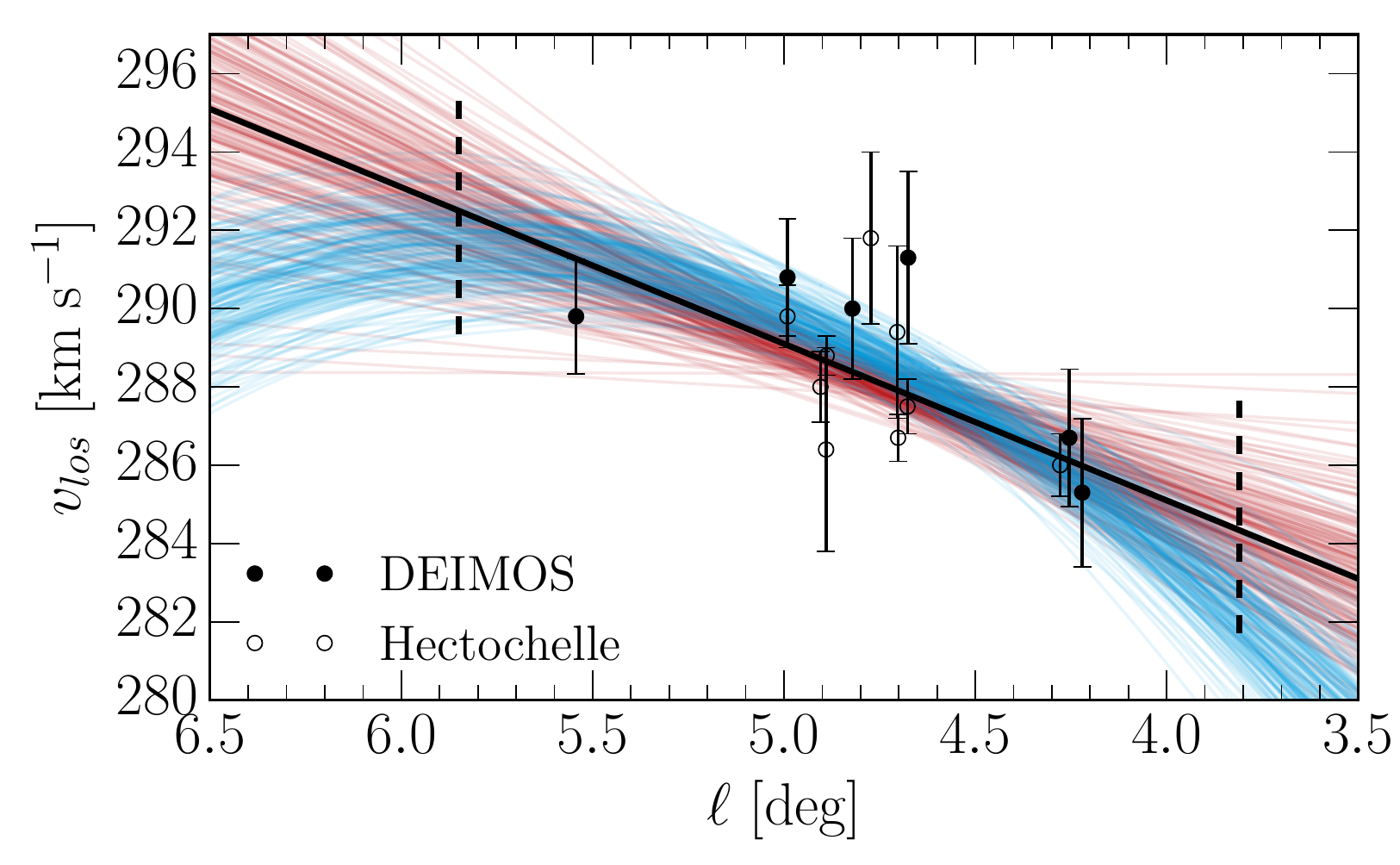}{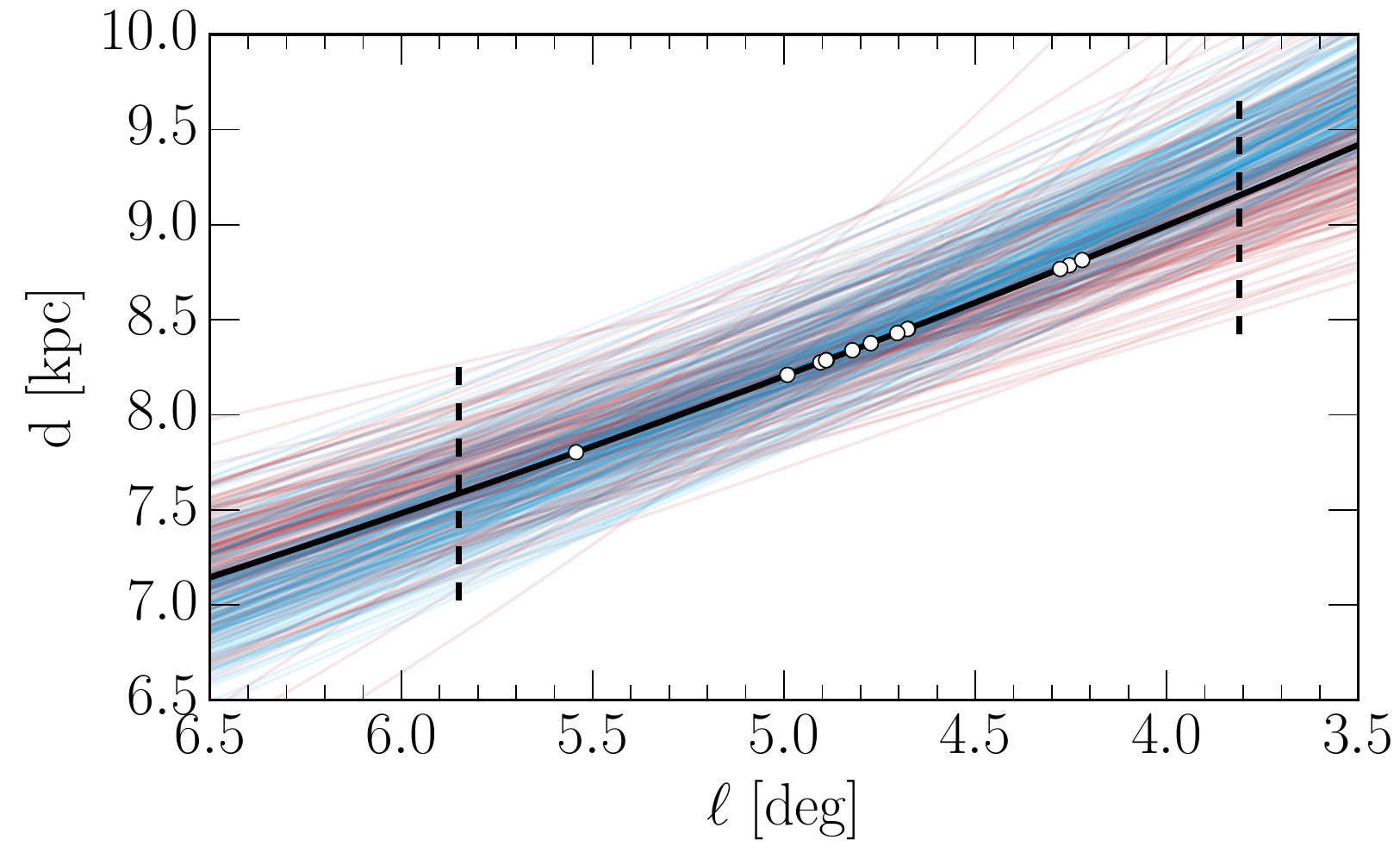}
\plottwo{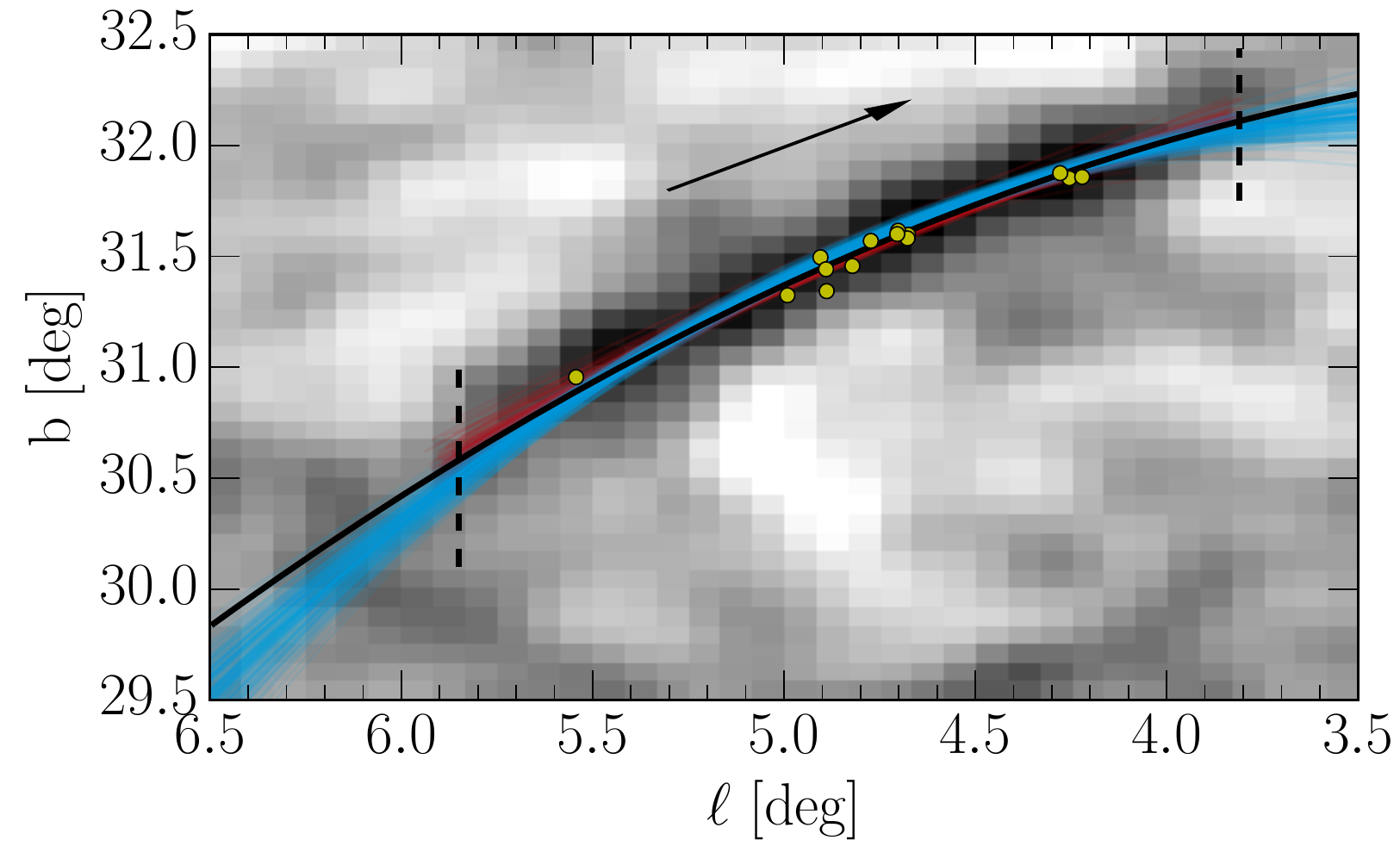}{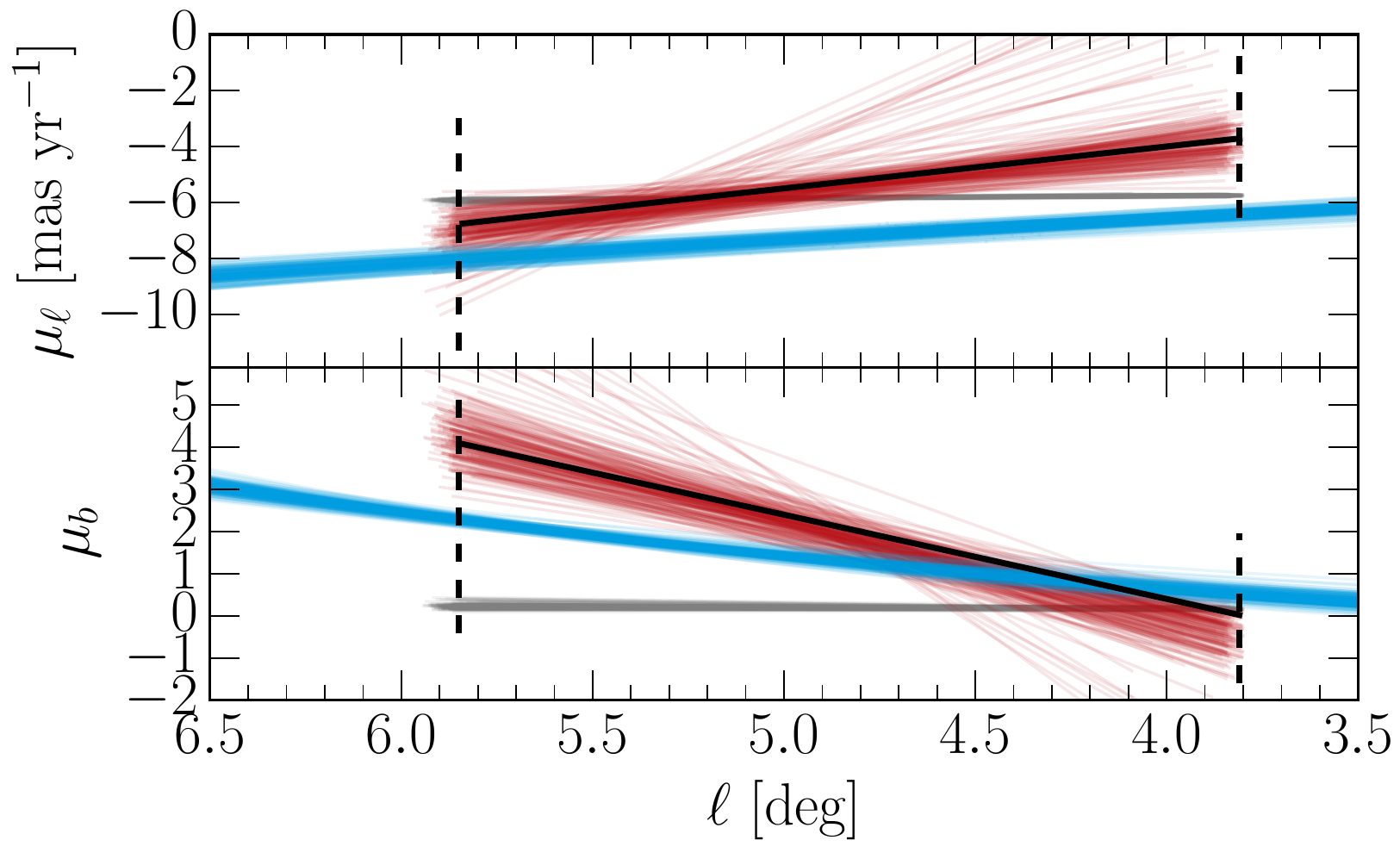}
\caption{
This plot compares line of sight velocities ({\em top left}), distances
({\em top right}), positions ({\em bottom left}), and proper motions
({\em bottom right}) calculated by \texttt{galpy} (thin blue lines) with models
derived from observations (thin red lines). The observed and calculated values
are consistent within uncertainties, with the exception of the proper motions,
where the model cannot match the apparent gradient of the proper motions along
the stream.
\label{galpy_orbit_comparison}}
\end{figure*}

The best-fit line of sight velocities, heliocentric distances, positions, 
and proper motions predicted by \texttt{galpy} orbits for each stream sample are
shown as thin semi-transparent blue lines in
Figure~\ref{galpy_orbit_comparison}. The maximum a posterior values, the median
and the central 68\% confidence intervals of orbital parameters are listed in
Table~\ref{stream_parameters}.

Overall, the observed and predicted mean values and gradients agree within
uncertainties. This agreement is not trivial. While there is always an orbit
that will fit a {\em single} star in some potential, the same is not true for a 
{\em stream} of stars. For example, given the observed gradients and mean values
in proper motion, distance, and position, the observed gradient in line of sight
velocity has to be positive, otherwise, there is a strong discrepancy with the
velocity predicted by the most probable orbit. Similarly, the observed gradient 
in distance modulus has to have a negative sign, otherwise a plausible orbit fit
cannot be achieved.

\begin{figure}
\plotone{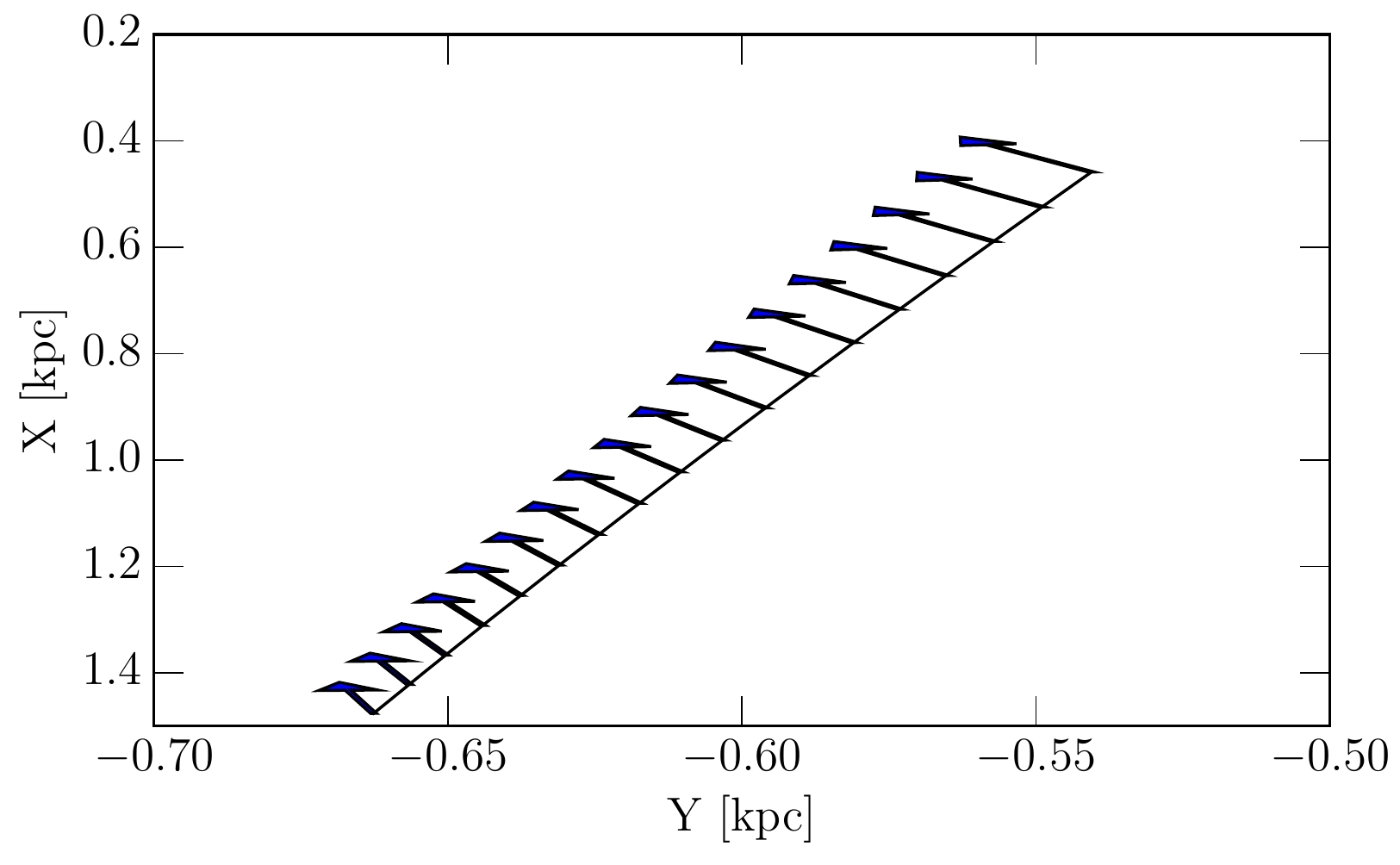}
\caption{
This plot illustrates the misalignment between the {\em observed} velocity
vectors of stream stars ({\em arrows}) and the extent of the stream ({\em solid 
line}) in the galactocentric Cartesian $X$ vs.~$Y$ plane (i.e., a top-down view 
of the Galactic plane). In this coordinate system, the Sun is at
$(X,Y,Z)=(8,0,0)$ kpc and the y-axis is positive toward galactic longitude
$l=270\arcdeg$.
\label{stream_missalign}}
\end{figure}

The most noticeable disagreement is between observed and predicted proper
motions (bottom right panel of Figure~\ref{galpy_orbit_comparison}), with the
observed proper motion in the longitude direction at $\ell_0=5\arcdeg$
($\overline{\mu_\ell}$) having a $-2.2$ mas yr$^{-1}$ offset with respect to the
proper motion predicted by the \texttt{galpy} orbit fit. The result of such
observed proper motion (i.e., $\overline{\mu_\ell}=-5.5$ mas yr$^{-1}$) is that
the velocity vectors of stream stars do not align with the extent of the stream
in the galactocentric $X$ vs.~$Y$ (and $Y$ vs.~$Z$) plane
(see Figure~\ref{stream_missalign}). The expected behavior would be for the
velocity vectors to be aligned with the extent of the stream (i.e., the stream
gets longer in the direction it is moving), which would happen for
$\overline{\mu_\ell}=-7.7$ mas yr$^{-1}$.

As we have already stated in Section~\ref{proper_motions}, candidate QSOs do not
show any statistically significant proper motion
($\mu^{QSO}_{\ell,b}=0.3\pm0.2$ mas yr$^{-1}$), and galaxies do not show any
bulk motion either. Thus, we have no indication that faulty proper motions are
the cause of this inconsistency between the velocity vector of the stream and
its extent.

We have repeated orbit fitting after adding a $-2.2$ mas yr$^{-1}$ offset to
proper motions in the longitude direction, and have found that our results do
not change. This was expected since we assume a 2 mas yr$^{-1}$ uncertainty in
proper motions when fitting orbits with \texttt{galpy}.

\begin{figure}
\plotone{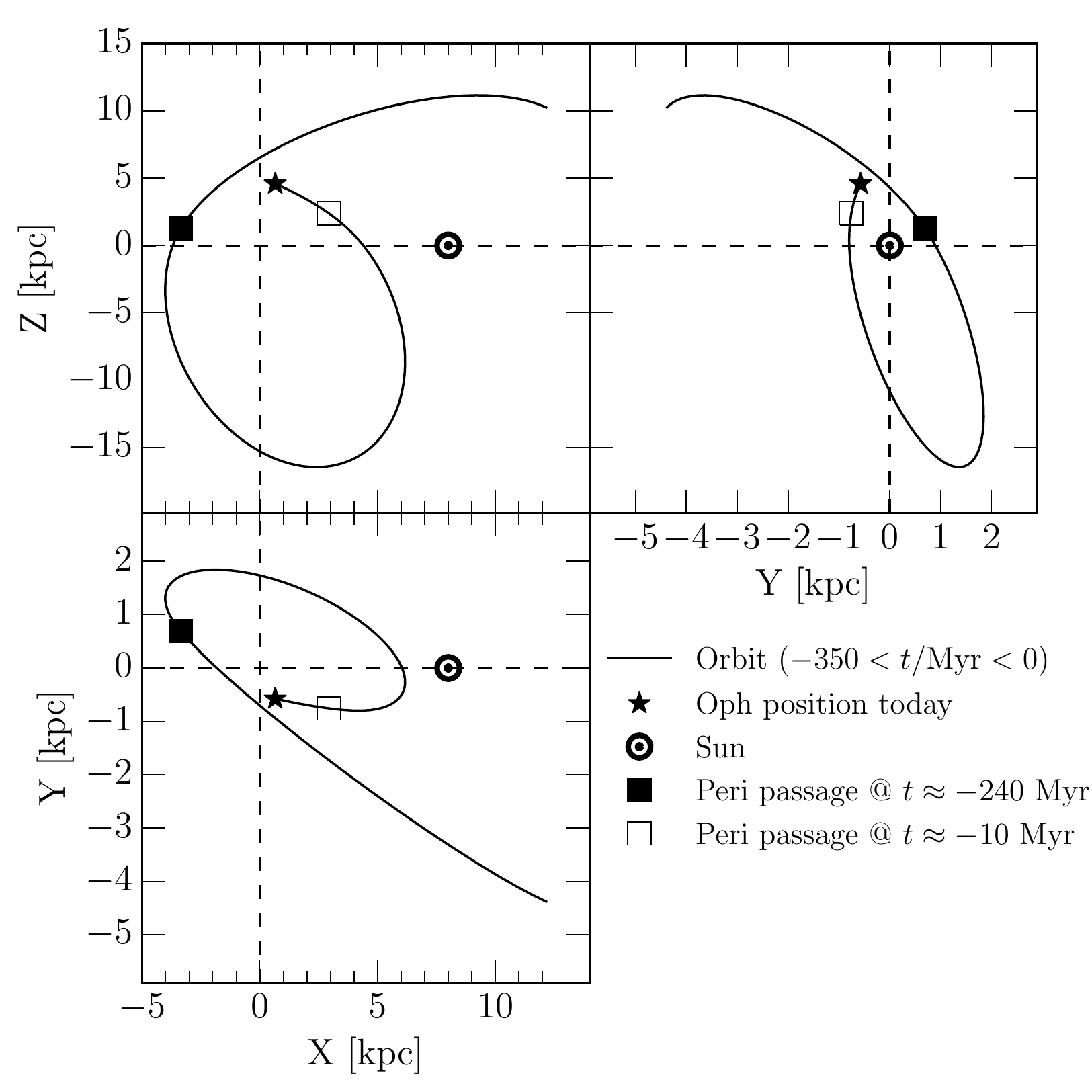}
\caption{
The orbit of the Ophiuchus stream in the past 350 Myr (about one orbital
period), shown in a right-handed galactocentric Cartesian coordinate system.
In this coordinate system, the Sun is at $(X,Y,Z)=(8,0,0)$ kpc and the y-axis is
positive toward galactic longitude $l=270\arcdeg$. Note the pericenter passage
at $t\approx-240$ Myr (solid square). Near this point in time, the stream was
also passing through the disk ($Z\sim0$ kpc, see the top left panel) and was 
experiencing strong tidal forces due to disk shocking (see
Figure~\ref{tidal_force}).
\label{orbit_XYZ}}
\end{figure}

Figure~\ref{orbit_XYZ} illustrates the orbit of the Ophiuchus stream in the past
350 Myr. We find that the stream has a relatively short orbital period of
$346_{-7}^{+11}$ Myr, and a fairly eccentric orbit ($e=0.65_{-0.01}^{+0.01}$),
with a pericenter of $3.57_{-0.06}^{+0.05}$ kpc and an apocenter of
$16.8_{-0.4}^{+0.6}$ kpc. About 10 Myr ago, the stream passed through its
pericenter and now it is moving away from the Galactic plane and towards the
Galactic center.

The above uncertainties in orbital parameters only account for the uncertainties
in position and velocity of the stream, and do not account for the uncertainty
in the distance of the Sun from the Galactic center ($R_0$), and the circular
velocity at the solar radius ($v_{circ}(R_0)$). To determine how the orbital
parameters change as a function of $R_0$ and $v_{circ}(R_0)$, we fit the orbit
of the stream assuming a Milky-Way-like potential fit to dynamical data as
described in Section 3.5 of \citet{bov15}, but assuming values of (8.5 kpc, 220
km s$^{-1}$), (8.0 kpc, 235 km s$^{-1}$), and (8.5 kpc, 235 km s$^{-1}$) for
($R_0, $$v_{circ}(R_0)$). When fitting these orbits, we model the stream using
the maximum a posterior values listed in Table~\ref{stream_parameters}.

We find that the best orbit fit is obtained for
$(R_0, v_{circ}(R_0))=(8.0, 220)$ (i.e., the default \texttt{galpy} values).
With respect to fiducial periods (i.e., those obtained assuming $R_0=8.0$ kpc
and $v_{circ}(R_0)=220$ km s$^{-1}$), modifying $R_0$ and $v_{circ}(R_0)$
changes periods of the Ophiuchus stream between $+2$ and $-75$ Myr. Other
orbital parameters do not change appreciably.

\section{Time of disruption}\label{disruption}

In the Introduction, we said that the short length of the Ophiuchus stream
suggests that its progenitor must have been disrupted fairly recently. As we
have shown in Section~\ref{proper_motion_extent}, part of the reason the stream
is so short {\em in projection}, is the viewing angle--we are observing the
stream almost end-on.

Even when the projection effects are taken into account, the deprojected length
is still fairly short, only 1.6 kpc. For comparison, the second shortest stellar
stream is the Pisces stream \citep[also known as the Triangulum stream,
\citealt{bon12}]{pisces} with a length of $\sim5.5$ kpc. Therefore, the length
of the stream still suggests that the stream formed recently, that is, it
suggests that the progenitor was recently disrupted.

As the progenitor of the Ophiuchus stream orbited the Galaxy, it would have
experienced the tidal force of the Galactic potential. This force can strip
stars from the progenitor and it could have been strong enough to completely
disrupt the progenitor. 

\begin{figure}
\plotone{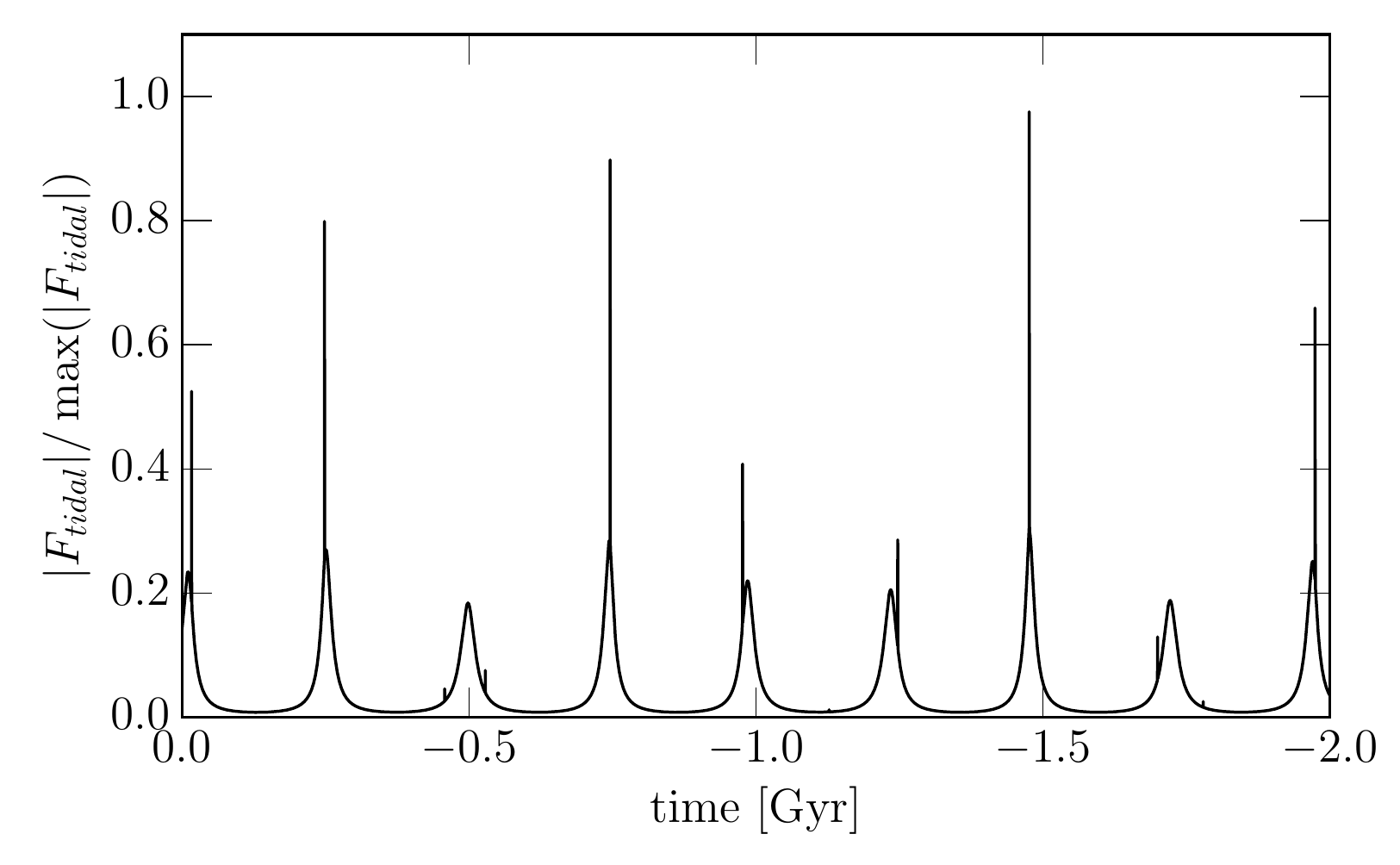}
\caption{
The tidal force acting on the Ophiuchus stream, normalized to the maximum tidal
force in the past 2 Gyr. The narrow peaks correspond to passages through the
disk (i.e., disk shocking, \citealt{osc72}) and the broader peaks correspond to
passages through the pericenter.
\label{tidal_force}}
\end{figure}

In order to examine the influence of the tidal force, we have calculated its
magnitude as a function of time for the most probable orbit of the Ophiuchus
stream. The magnitude of the tidal force was calculated by finding the largest
eigenvalue of the following matrix
\begin{eqnarray}
J & = &
\left(
\begin{array}{cc}
    \frac{d^2\Phi}{dR^2} & \frac{d^2\Phi}{dRdZ} \\
    \frac{d^2\Phi}{dZdR} & \frac{d^2\Phi}{dZ^2}
\end{array}
\right),
\end{eqnarray}
where $\Phi$ is value of the \texttt{galpy} \texttt{MWPotential2014} potential
at the position of the progenitor, and $R$ and $Z$ are coordinates in the
cylindrical galactocentric system. The result is shown in
Figure~\ref{tidal_force}.

We find that the tidal force is the strongest during pericenter+disk passages,
and that the progenitor of the stream could have been disrupted during one of
those passages. To find if the progenitor could plausibly have been disrupted
during one of these passages, we can use \texttt{galpy}.

\begin{figure}
\plotone{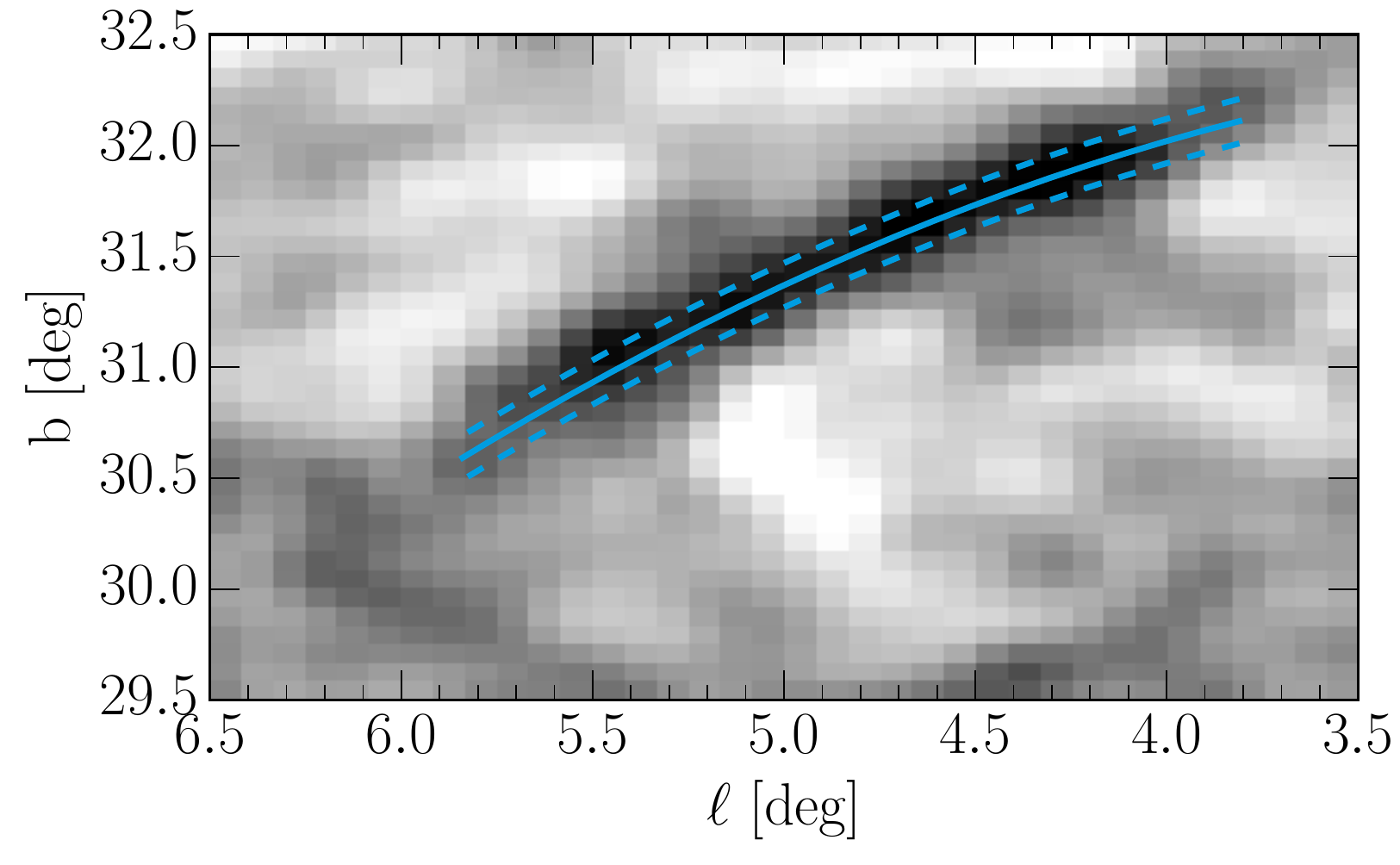}

\plotone{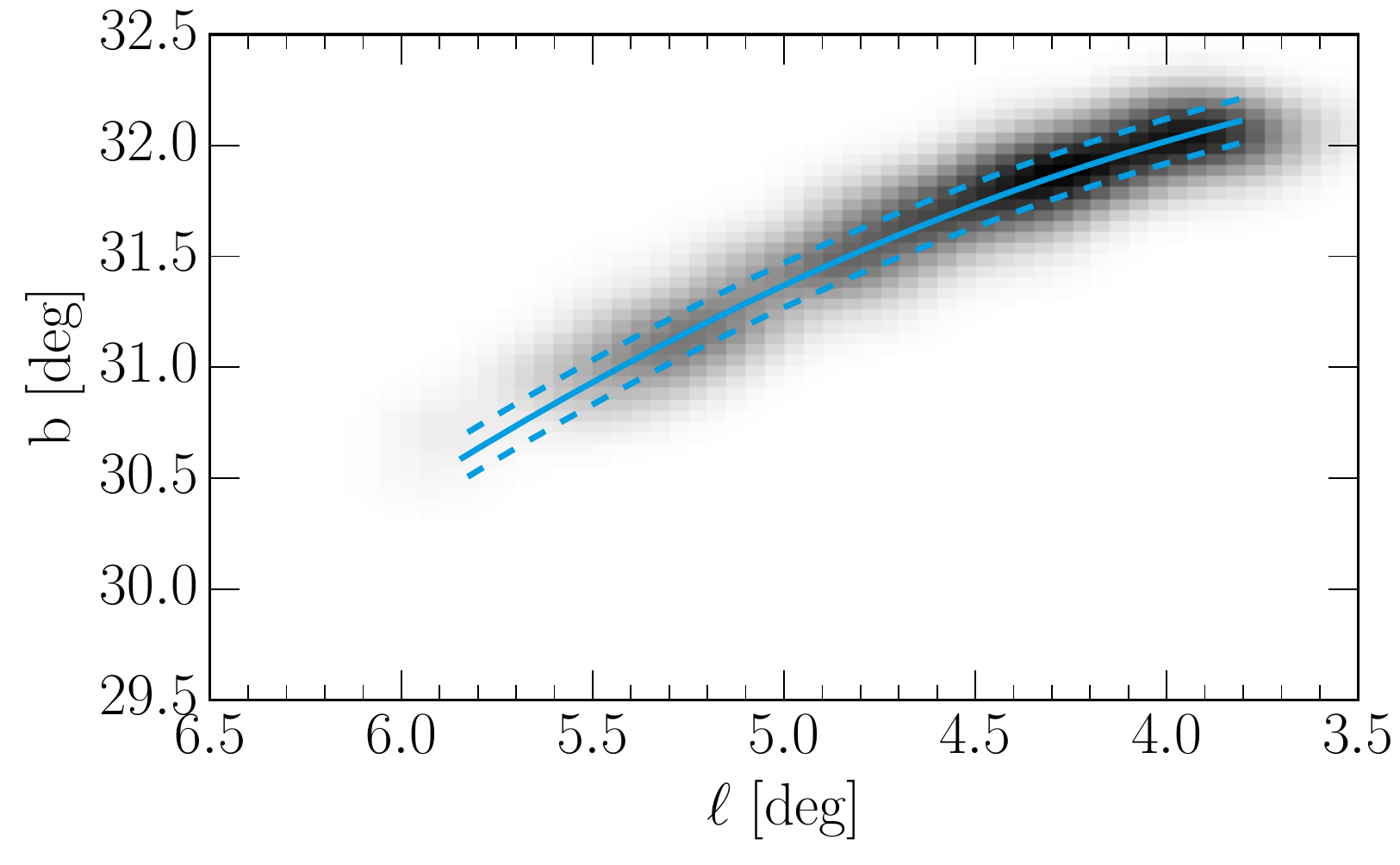}

\plotone{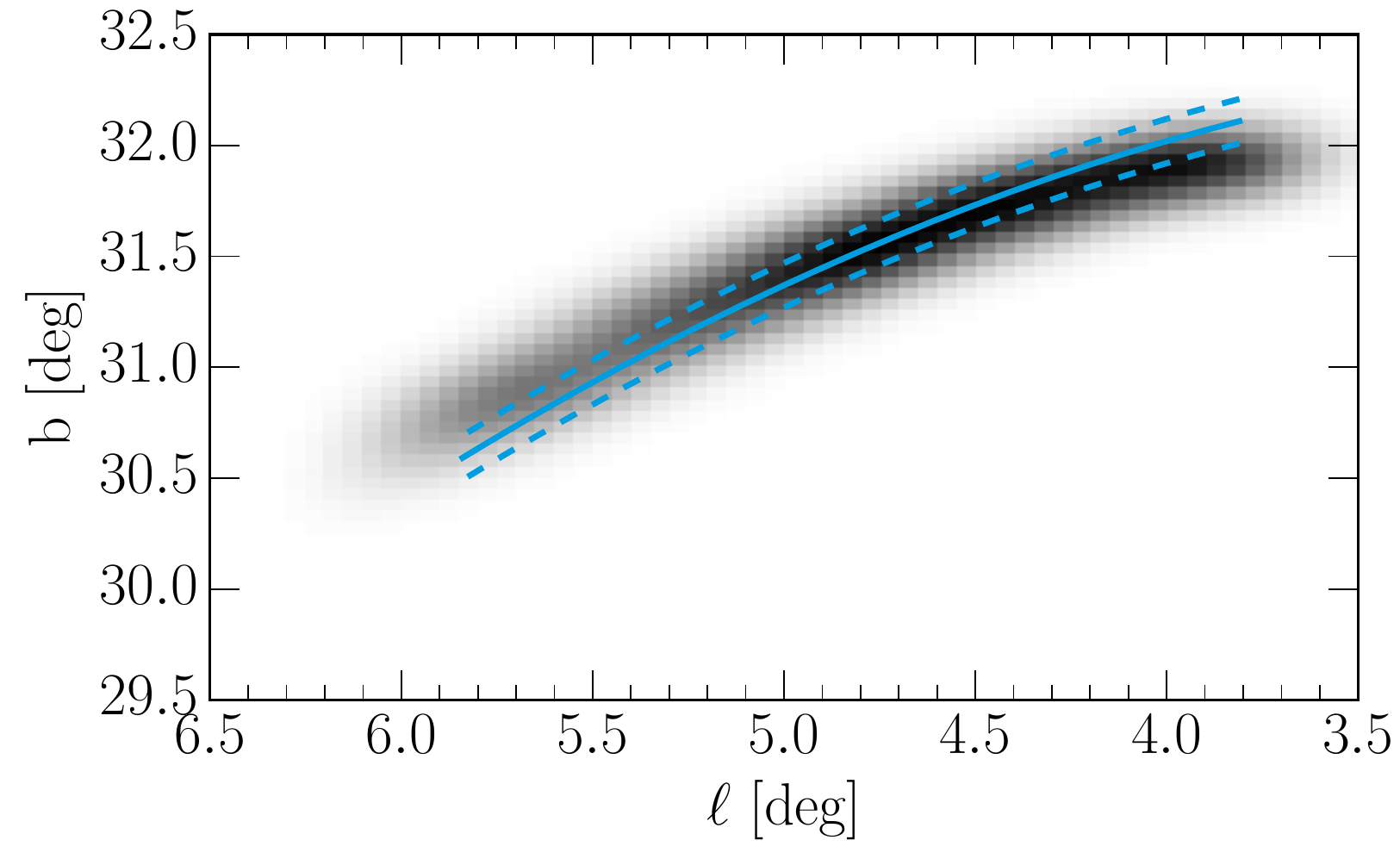}
\caption{
A comparison of the observed number density map of the stream ({\em top}), a map
created from a mock stream generated using \texttt{galpy} ({\em middle}), and a 
map created from a $N$-body stream generated using NEMO. In all panels, the
solid line shows the most probable position of the stream and the dashed lines
illustrate its $1\sigma$ width. The mock \texttt{galpy} stream was generated
assuming time of disruption $t_{dis}=170$ Myr, and velocity dispersion
$\sigma_{v}=0.4$ km s$^{-1}$. Note a good agreement between the length and width
of the observed, \texttt{galpy} and NEMO streams.
\label{observed_vs_mock_comparison}}
\end{figure}

Given an orbit, the time of disruption $t_{dis}$, and the velocity dispersion of
the progenitor $\sigma_v$, \texttt{galpy} can generate a mock stream using the
modeling framework of \citet{bov14}. For a fixed $\sigma_v$, the disruption
time, $t_{dis}$, is proportional to the stream's length (i.e., older streams are
longer). In Section~\ref{vlos}, we measured $s=0.4$ km s$^{-1}$ as the median
velocity dispersion of the stream. We find that by setting $\sigma_v$ to the
same value, the \texttt{galpy} mock stream has a velocity dispersion
$\tilde{s} \approx 0.4$ km s$^{-1} = s$. We further find that $t_{dis}\sim170$
Myr provides a good match between the observed and mock streams (see
Figure~\ref{observed_vs_mock_comparison}). While we do not perform an exhaustive
search of the $(\sigma_v,t_{dis})$ parameter space, very different values of
$\sigma_v$ or $t_{dis}$ produce values of $\tilde{s}$ or a total length that are
not consistent with observations. This result, and the fact that there was a
disk+pericenter passage 240 Myr ago, strongly suggest that the stream formed
about 240 Myr ago (i.e., that the progenitor was disrupted at that time).

As shown by \citet{joh98}, the velocity dispersion of the progenitor scales with
the mass of the progenitor as $\sigma_v \propto M_{dyn}^{1/3}$. In
\texttt{galpy},  the model used for stream generation was calibrated using a
progenitor of mass $M_{dyn}=2\times10^4$ $M_\sun$. The stream formed by the
tidal disruption of this progenitor could be modeled with $\sigma_v=0.365$ km
s$^{-1}$ \citep{bov14}. Using the above scaling relation and $\sigma_v=0.4$ km
s$^{-1}$, we find that the progenitor of the Ophiuchus stream had a mass of
$M_{dyn}\sim2\times10^4$ $M_\sun$.

It is important to note that \texttt{galpy} creates only the stream, and that
stars associated with the progenitor are not part of the mock stream (i.e., are 
not shown in the middle panel of Figure~\ref{observed_vs_mock_comparison}). This
is the reason an overdensity of stars (i.e., the progenitor) is not visible in
the mock \texttt{galpy} stream.

To create a more realistic stream that includes a progenitor, we use the
gyrfalcON code \citep{deh00,deh02} in the NEMO toolkit \citep{teu95}. We set up
the progenitor as a King cluster \citep{kin66} with a mass of $1\times10^4$
$M_\sun$, tidal radius of 94 pc and a ratio of the central potential to the
velocity dispersion squared of 2.0. The cluster is sampled using 20,000
particles and is evolved for $\sim365$ Myr in the \texttt{MWPotential2014}
potential. The initial conditions are those at the $t\sim-365$ Myr apocenter of 
the most probable orbit of the Ophiuchus stream.

The number density map of the $N$-body stream created using gyrfalcON is shown
in the bottom panel of Figure~\ref{observed_vs_mock_comparison}. The width and
the length of the $N$-body stream match the observed stream fairly well. For
comparison, if the cluster is evolved starting from the apocenter at
$t\sim-870$ Myr, the resulting stream is much longer and inconsistent with
observations. Based on this more realistic simulation, we conclude that the
Ophiuchus stream likely formed about 240 Myr ago and that its progenitor was
fully disrupted during a single disk+pericenter passage.

\section{Conclusions and Summary}\label{conclusions}

In this paper, we have presented follow-up spectroscopy and an astrometric and
photometric analysis of the Ophiuchus stellar stream in the Milky Way, recently
discovered by \citet{ber14b} in PS1 data. We have been able to put together a
comprehensive, empirical description of the Ophiuchus stream in phase space: we
succeeded in determining the mean phase-space coordinates in all six dimensions,
along with the gradients of those coordinates along the stream (see
Table~\ref{stream_parameters} for a summary of stream parameters).

Overall, phase-space data along the stream can be well matched by an orbit in a
fiducial Milky Way potential: Ophiuchus is truly a thin and long ($\sim1.6$~kpc)
stellar stream, 50 times longer than wide, that appears $6:1$ foreshortened in
projection; it is on a highly inclined orbit with only a 350 Myr orbital period;
it is receding from us at nearly 300 km s$^{-1}$ and has just passed its
pericenter at $\sim3$ kpc from the Galactic center. This makes Ophiuchus the
innermost stellar stream known in our Galaxy. It is also the only known
kinematically cold stellar stream to be seen nearly end-on.

The homogeneously metal-poor (${\rm [Fe/H]}=-2.0$ dex), $\alpha$-enhanced
(${\rm [\alpha/Fe]\sim0.4}$ dex) and old stellar population ($\sim12$ Gyr old),
and the small line of sight velocity dispersion we found ($<1$ km s$^{-1}$),
strongly confirm the notion that the progenitor of the stream must have been a
globular cluster. If the detected part of the stream encompasses most of the
progenitor's stars, then the progenitor's tidal radius was $\sim90$ pc and its
mass was $\sim2\times10^4$ $M_\sun$ (certainly greater than $\sim7\times10^3$
$M_\sun$). In this respect, the Ophiuchus and the GD-1 stream \citep{gd06,kop10}
can be considered identical twins, as they have the same metallicity, the same
mass, and no detectable progenitors.

Our analysis, however, leaves a number of questions open. First, the most
probable orbit in the fiducial potential is not quite able to match the proper
motions and their gradients along the stream. A thorough exploration whether
there are axiymmetric or non-axisymmetric potentials that might be able to
remedy this tension remains to be done. Alternatively, this discrepancy may
indicate a problem with the proper motion data. While we have done our best to
obtain good proper motions, we cannot fully dismiss this possibility. However,
we expect that the proper motions provided by the the ongoing GAIA mission
\citep{per01} will resolve this discrepancy in the near future.

Second, the present analysis does not yet use the stream phase-space data to
provide new constraints on the Galactic potential. In principle, the Ophiuchus
stream can provide constraints on the Galactic potential at about 4 kpc above
the Galactic center, a location where few other constraints exist. As the top
left panel of Figure~\ref{galpy_orbit_comparison} shows, the line of sight
velocity of the stream is predicted to decrease to about 282 km s$^{-1}$ at
galactic longitude $l=4\arcdeg$. By measuring line of sight velocities of stars
at this position, we can identify members of the Ophiuchus stream and test the
assumed potential. Similarly, the potential may also be tested by identifying
stream members at $l\ga5.8\arcdeg$, where the stream is predicted to curve
toward the galactic latitude $b=30.5\arcdeg$ (see the blue lines in the bottom
left panel of Figure~\ref{galpy_orbit_comparison}).

And finally, our analysis suggests that the progenitor of the stream was fully
disrupted during a single disk+pericenter passage about 240 Myr ago. As $N$-body
simulations show, this disruption was strong enough to smooth out the
distribution of stars along the stream and effectively erase all evidence of the
progenitor. If this scenario is correct, then the answer to the question ``Where
is the progenitor of the Ophiuchus stream?'' is fairly simple--the Ophiuchus
stream is all that is left of the progenitor.

While the above scenario seems to answer one question, it creates another one.
The fact that the $N$-body stream ends up being too long if the cluster is
evolved for more than $\sim400$ Myr, suggests that the progenitor could not have
been on the current orbit for more than $\sim400$ Myr. If that is true, how did 
the progenitor end up on the current orbit and what was its original orbit? We
expect that detailed $N$-body simulations that include the interactions with the
Galactic bar will provide a more definitive answer to these questions and plan
to pursue this approach.

Our finding that the progenitor could not have been on the current orbit for
more than $\sim400$ Myr is based on the comparison of the length of the observed
and the $N$-body stream. However, what if the stream is actually much longer,
but is simply not observed as such in current data? This could happen if, for
example, the stream suddenly fans out and thus its surface brightness drops
below our detection limit. The stream may fan out due to interactions with dark
matter subhalos (bottom left panel of Figure~3 by \citealt{bon14}), due to being
on a chaotic orbit (bottom panels of Figure~11 by \citealt{far15}; also
\citealt{apw15}), or due to being in a triaxial potential (Figure~4 by
\citealt{pea15}). A longer stream would imply that the progenitor has been
undergoing disruption for a longer time, which would make the observed orbit
more plausible as the progenitor's original orbit (i.e., a change in orbit would
not be necessary).

The true extent of the stream may be constrained by identifying stream members
along the predicted extent of the stream or in a fan-out pattern, via
line of sight velocities. As Figure~\ref{rv_hist} shows, the stream's high
velocity makes the separation of member stars from field stars an easy task. We 
have already started a follow-up spectroscopic program with the goal of
identifying additional stream members, and hope to better constrain the length, 
orbit, and possible fanning-out of the Ophiuchus stream in the near future.

\acknowledgments

B.S.~acknowledges funding from the European Research Council under the European
Union’s Seventh Framework Programme (FP 7) ERC Grant Agreement
n.~${\rm [321035]}$. C.I.J.~gratefully acknowledges support from the Clay
Fellowship, administered by the Smithsonian Astrophysical Observatory. A.P.W.~is
supported by a National Science Foundation Graduate Research Fellowship under
Grant No.~11-44155. The Pan-STARRS1 Surveys (PS1) have been made possible
through contributions by the Institute for Astronomy, the University of Hawaii, 
the Pan-STARRS Project Office, the Max-Planck Society and its participating
institutes, the Max Planck Institute for Astronomy, Heidelberg and the Max
Planck Institute for Extraterrestrial Physics, Garching, The Johns Hopkins
University, Durham University, the University of Edinburgh, the Queen's
University Belfast, the Harvard-Smithsonian Center for Astrophysics, the Las
Cumbres Observatory Global Telescope Network Incorporated, the National Central 
University of Taiwan, the Space Telescope Science Institute, and the National
Aeronautics and Space Administration under Grant No.~NNX08AR22G issued through
the Planetary Science Division of the NASA Science Mission Directorate, the
National Science Foundation Grant No.~AST-1238877, the University of Maryland,
Eotvos Lorand University (ELTE), and the Los Alamos National Laboratory. Some of
the data presented herein were obtained at the W.M.~Keck Observatory, which is
operated as a scientific partnership among the California Institute of
Technology, the University of California and the National Aeronautics and Space 
Administration. The Observatory was made possible by the generous financial
support of the W.M.~Keck Foundation. The authors wish to recognize and
acknowledge the very significant cultural role and reverence that the summit of 
Mauna Kea has always had within the indigenous Hawaiian community. We are most
fortunate to have the opportunity to conduct observations from this mountain.
Observations reported here were obtained at the MMT Observatory, a joint
facility of the Smithsonian Institution and the University of Arizona.

{\it Facilities:} \facility{PS1}, \facility{Keck:I (DEIMOS)}, \facility{MMT (Hectochelle)}

\bibliographystyle{apj}
\bibliography{ms}

\newpage

\begin{turnpage}
\begin{deluxetable*}{lrrrrrrrrrrr}
\tabletypesize{\scriptsize}
\setlength{\tabcolsep}{0.02in}
\tablecolumns{12}
\tablewidth{0pc}
\tablecaption{Ophiuchus Stream Member Stars\label{table1}}
\tablehead{
\colhead{Name} & \colhead{R.A.} & \colhead{Decl.} &
\colhead{$g_{P1}$} & \colhead{$r_{P1}$} & \colhead{$i_{P1}$} &
\colhead{$z_{P1}$} & \colhead{$y_{P1}$} &
\colhead{$v_{los}$} & \colhead{DM} &
\colhead{$\mu_l$} & \colhead{$\mu_b$} \\
\colhead{ } & \colhead{(deg)} & \colhead{(deg)} &
\colhead{(mag)} & \colhead{(mag)} & \colhead{(mag)} &
\colhead{(mag)} & \colhead{(mag)} &
\colhead{(km s$^{-1}$)} & \colhead{(mag)} &
\colhead{(mas yr$^{-1}$)} & \colhead{(mas yr$^{-1}$)}
}
\startdata
bhb1 & 241.52271 & -7.01555 & $16.05\pm0.02$ & $16.11\pm0.02$ & $16.22\pm0.01$ & $16.25\pm0.02$ & $16.25\pm0.02$ & $286.7\pm1.8$ & $14.72_{-0.06}^{+0.06}$ & $-4.1\pm2.1$ & $2.6\pm2.1$ \\
bhb2 & 241.49994 & -7.03409 & $16.02\pm0.02$ & $16.09\pm0.02$ & $16.21\pm0.02$ & $16.25\pm0.02$ & $16.23\pm0.02$ & $285.3\pm1.9$ & $14.73_{-0.06}^{+0.06}$ & $-4.4\pm2.2$ & $2.1\pm2.2$ \\
bhb3 & 242.13551 & -6.87785 & $15.96\pm0.02$ & $15.99\pm0.01$ & $16.11\pm0.02$ & $16.16\pm0.02$ & $16.13\pm0.02$ & $290.0\pm1.8$ & $14.61_{-0.05}^{+0.05}$ & $-5.2\pm2.0$ & $2.4\pm2.0$ \\
bhb4 & 241.94714 & -6.88995 & $16.22\pm0.02$ & $16.32\pm0.01$ & $16.48\pm0.02$ & $16.54\pm0.02$ & $16.56\pm0.02$ & $291.3\pm2.2$ & $14.64_{-0.06}^{+0.05}$ & $-5.1\pm2.2$ & $3.3\pm2.1$ \\
bhb6 & 242.33018 & -6.84405 & $15.67\pm0.01$ & $15.64\pm0.01$ & $15.59\pm0.01$ & $15.60\pm0.02$ & $15.54\pm0.02$ & $290.8\pm1.5$ & $14.58_{-0.05}^{+0.05}$ & $-5.6\pm1.3$ & $5.2\pm1.4$ \\
bhb7 & 242.91469 & -6.69329 & $15.66\pm0.01$ & $15.56\pm0.02$ & $15.57\pm0.01$ & $15.54\pm0.02$ & $15.50\pm0.02$ & $289.8\pm1.5$ & $14.47_{-0.06}^{+0.05}$ & $-7.4\pm1.5$ & $7.1\pm1.7$ \\
\hline
bhb6 & & & & & & & & $289.8\pm0.8$ & & & \\
rgb1 & 241.51689 & -6.98511 & $17.71\pm0.02$ & $17.18\pm0.02$ & $16.91\pm0.02$ & $16.78\pm0.02$ & $16.71\pm0.02$ & $286.0\pm0.8$ & $14.72_{-0.06}^{+0.06}$ & $-6.2\pm1.5$ & $0.7\pm1.4$ \\
rgb2 & 241.94649 & -6.86113 & $17.34\pm0.02$ & $16.74\pm0.02$ & $16.46\pm0.02$ & $16.32\pm0.02$ & $16.25\pm0.02$ & $286.7\pm0.6$ & $14.64_{-0.05}^{+0.05}$ & $-4.5\pm1.4$ & $1.1\pm1.4$ \\
rgb3 & 241.96089 & -6.89873 & $17.64\pm0.02$ & $17.08\pm0.02$ & $16.83\pm0.02$ & $16.68\pm0.02$ & $16.62\pm0.02$ & $287.5\pm0.7$ & $14.64_{-0.06}^{+0.05}$ & $-6.5\pm1.5$ & $3.0\pm1.4$ \\
rgb4 & 242.26139 & -6.90190 & $17.05\pm0.02$ & $16.45\pm0.02$ & $16.17\pm0.01$ & $16.02\pm0.02$ & $15.93\pm0.02$ & $288.8\pm0.5$ & $14.60_{-0.05}^{+0.05}$ & $-23.7\pm1.4$ & $-14.8\pm1.5$ \\
rgb5 & 242.14832 & -6.79765 & $17.79\pm0.02$ & $17.23\pm0.02$ & $16.96\pm0.02$ & $16.82\pm0.02$ & $16.74\pm0.02$ & $288.0\pm0.9$ & $14.60_{-0.05}^{+0.05}$ & $-6.0\pm1.3$ & $2.4\pm1.4$ \\
sgb1 & 241.95962 & -6.86881 & $18.90\pm0.02$ & $18.53\pm0.02$ & $18.38\pm0.02$ & $18.31\pm0.02$ & $18.28\pm0.02$ & $289.4\pm2.2$ & $14.63_{-0.05}^{+0.05}$ & $-5.8\pm2.0$ & $1.2\pm2.1$ \\
msto1 & 242.02040 & -6.84122 & $19.22\pm0.02$ & $18.89\pm0.02$ & $18.74\pm0.02$ & $18.69\pm0.02$ & $18.62\pm0.03$ & $291.8\pm2.2$ & $14.62_{-0.05}^{+0.05}$ & $-6.9\pm2.4$ & $4.4\pm2.4$ \\
msto2 & 242.18360 & -6.84056 & $19.10\pm0.02$ & $18.70\pm0.02$ & $18.54\pm0.02$ & $18.45\pm0.02$ & $18.42\pm0.03$ & $286.4\pm2.6$ & $14.60_{-0.05}^{+0.05}$ & $-3.4\pm2.3$ & $2.3\pm2.2$
\enddata
\tablecomments{The horizontal line separates stars observed by DEIMOS and
Hectochelle. The name indicates the likely evolutionary stage inferred from
isochrone fitting. The PS1 photometry is {\em not} corrected for extinction. The
uncertainty in $v_{los}$ includes the uncertainty from cross-correlation/fitting
and the uncertainty in the zero-point of wavelength calibration. The DM
indicates the average DM of the stream at the position of the star, and
the uncertainties are $68\%$ confidence intervals. The proper motion in the
galactic longitude direction, $\mu_l$, includes the $\cos b$ term.}
\end{deluxetable*}
\end{turnpage}
\clearpage
\global\pdfpageattr\expandafter{\the\pdfpageattr/Rotate 90}

\end{document}